\documentclass[twocolumn
]{aastex631}
\pdfoutput=1

\usepackage{CJK}
\usepackage{xspace}
\usepackage{booktabs}
\usepackage{lmodern}
\usepackage{slantsc}
\usepackage{comment}
\usepackage{longtable}
\usepackage{url}

\newcommand{\teff}{$T_{\mathrm{eff}}$}
\newcommand{\logg}{$\log(g)$}
\newcommand{\msun}{M$_\odot$} 

\received{August 10, 2026}
\revised{September 17, 2026}
\accepted{September 18, 2026}

\shorttitle{Gaia Massive Metal-Polluted White Dwarfs}
\shortauthors{Ould Rouis et al.}

\graphicspath{{./}{figures/}}

\begin{document}

\begin{CJK*}{UTF8}{gbsn}

\title{Searching for Massive Remnant Planetary System Hosts with White Dwarf Gaia XP Spectra}


\author[0009-0002-6065-3292]{Lou~Baya~Ould~Rouis}
\email{lbor@bu.edu}
\affiliation{Department of Astronomy \& Institute for Astrophysical Research Boston University Boston, MA 02215, USA}

\author[0000-0001-5941-2286]{J.~J.~Hermes}
\affiliation{Department of Astronomy \& Institute for Astrophysical Research Boston University Boston, MA 02215, USA}

\author[0000-0003-2368-345X]{Pierre~Bergeron}
\affiliation{D\' epartement de Physique, Universit\' e\ de Montr\' eal, C.P. 6128, Succ.~Centre-Ville, Montr\' eal, Qu\' ebec, Canada}

\author[0000-0002-9632-1436]{Simon Blouin}
\affiliation{Department of Physics and Astronomy, University of Victoria, Victoria, BC V8W 2Y2, Canada}

\author[0000-0002-6270-8624]{Stefan~Arseneau}
\affiliation{Department of Astronomy \& Institute for Astrophysical Research Boston University Boston, MA 02215, USA}

\author[0000-0001-7296-3533]{Tim~Cunningham}
\affiliation{Center for Astrophysics | Harvard \& Smithsonian, 60 Garden St., Cambridge, MA 02138, USA}

\author[0000-0001-9632-7347]{Joseph~A.~Guidry}
\affiliation{Department of Astronomy \& Institute for Astrophysical Research Boston University Boston, MA 02215, USA}

\author[0009-0000-8806-5623]{Ciarra~Coston}
\affiliation{Department of Astronomy \& Institute for Astrophysical Research Boston University Boston, MA 02215, USA}


\begin{abstract}

Remnant planetary systems around massive white dwarfs ($M_{\rm WD} > 0.8\,{\rm M}_\odot$) are intrinsically rare and remain poorly characterized. As the evolutionary endpoints of intermediate-mass B- and A-type stars, these white dwarfs provide valuable constraints on the formation and survival of planetary systems around higher-mass progenitors. In the absence of large, dedicated spectroscopic surveys, extremely-low-resolution Gaia XP spectra provide a powerful means of identifying candidate metal-polluted white dwarfs.
We obtained follow-up observations with low-resolution optical spectroscopy (R=500--4000) to search primarily for Ca~{\sc ii}~H\,\&\,K absorption in massive, metal-polluted white dwarf candidates selected from Gaia XP spectra to increase the sample of massive remnant planetary system hosts among objects with effective temperatures below $\sim15{,}000$ K. We find that Gaia XP spectra recover metal-polluted white dwarfs efficiently, confirming 82$\pm$5\% recovery of photospheric metals (59 of 72 targets) in  high-confidence candidates of our sample (with classifier probability $P_{\rm DZ} > 0.65$). However, Gaia photometric masses derived from $G$, $G_{BP}$, and $G_{RP}$ broad-band photometry are not sufficient to identify the most massive systems, as reliable mass determinations require detailed metal-enriched atmospheric fits. 
We present a selection of cool white dwarfs confirmed to be both massive and metal polluted, demonstrating that Gaia XP spectra can substantially expand the sample of white dwarf exoplanet hosts. Our search led to the discovery of a new transiting debris system, SDSS\,J1624+5404, the 21st published system to date and eighth to show periodic transits. 

\end{abstract} 
\vspace{-5mm}


\section{Introduction} \label{sec:intro}

White dwarfs are the endpoint of stellar evolution for progenitors with main sequence masses lower than 8$-$10\,$\rm M_{\odot}$. As exoplanet demographic studies become common \citep{2012Natur.481..167C, 2021AcA....71....1P}, white dwarfs provide a unique means of probing planetary systems around intermediate-mass stars that are poorly accessible to traditional exoplanet detection techniques. Radial velocity surveys are less sensitive to higher-mass stars because they typically exhibit fewer narrow spectral lines, more rapid rotation, and greater photospheric variability (e.g., \citealt{2002AN....323..392H}). Transit surveys are likewise less effective, as transit depths scale inversely with stellar radius, making them intrinsically more sensitive to planets orbiting smaller stars than to those around more massive hosts.

Planets on distant orbits are expected to survive to the white dwarf stage \citep{2012ApJ...761..121M}. The presence of remnant planetary systems around white dwarfs is commonly inferred through atmospheric metal pollution \citep{2003ApJ...596..477Z, 2005A&A...432.1025K}, produced by the accretion of scattered tidally disrupted planetesimals onto the star (e.g. \citealt{2016NewAR..71....9F, 2016RSOS....350571V, 2022Natur.602..219C}). The strong surface gravity of white dwarfs causes heavy elements to sink below the photosphere on short timescales, so the detection of metals in white dwarf spectra requires ongoing or recent accretion \citep{1979ApJ...231..826F, 1986ApJS...61..197P, 1993ApJS...87..345D, 2006A&A...453.1051K, 2009A&A...498..517K}. In white dwarfs with \teff\,$\lesssim$ 10,000\,K, deep convection zones increase mixing and extend diffusion timescales, but do not alter the requirement for photospheric metals to be externally accreted \citep{2009A&A...498..517K, 2019ApJ...872...96B, 2019MNRAS.488.2503C, 2025MNRAS.544.2098B}. In optical spectra, the Ca~{\sc ii}~H\,\&\,K absorption lines are prominent tracers of this process \citep{2005A&A...432.1025K, 2007ApJ...663.1291D, 2018MNRAS.477...93H, 2019ApJ...885...74C}, making the measurement of photospheric metals an observational window into the long-term survival and evolution of planetary systems. 

A substantial fraction of white dwarfs are expected to host remnant planetary systems, as over 40\% of white dwarfs exhibit metal pollution (e.g., \citealt{2014A&A...566A..34K, 2019MNRAS.487..133W}), or roughly 11\% from occurrences in optical spectroscopy alone \citep{2024MNRAS.527.8687O}. However, this fraction is 4-$\sigma$ lower among massive white dwarfs ($M > 0.8\,\rm M_{\odot}$; \citealt{2024ApJ...976..156O}). The lack of massive metal-polluted white dwarfs likely results from conditions during planetary formation or evolution rather than stellar interactions, and cannot be explained by atomic diffusion and thermohaline mixing alone \citep{2026A&A...707A.130D}. 
Despite nearly 1750 known metal-polluted white dwarfs as of February~2026 per the Planetary Enriched White Dwarf Database (PEWDD; \citealt{2024A&A...691A.352W}), the lack of homogeneous mass estimates limits the overall understanding of the demographics and evolutionary histories of planetary systems around more massive progenitors. 

It is essential to expand the sample of well-characterized massive ($M_{\rm  WD} > 0.8\,\rm M_{\odot}$), metal-polluted white dwarfs, which originate from stars with masses greater than $3.5\,\rm M_{\odot}$ on the main sequence, in order to determine what conditions are required for planetary formation and survival. In particular, finding the most massive metal-polluted white dwarfs places constraints on which stars can form and retain planets, which in turn constrains the short disk lifetimes of massive stars \citep{2009ApJ...690.1539G}. 

The Gaia mission \citep{2016A&A...595A...1G} has revolutionized our understanding of white dwarfs by identifying ${\simeq} 359{,}000$ high-probability white dwarf candidates across the Galaxy \citep{2021MNRAS.508.3877G}. In addition, Gaia extremely low-resolution XP spectra (R$\sim$50) enable broad spectral classification \citep{2021A&A...652A..86C, 2023A&A...674A..33G}. Rather than providing conventional flux-calibrated spectra, the Gaia archive stores XP observations as coefficients of basis functions describing the BP/RP dispersed spectra, that can be reconstructed into spectra spanning 3300$-$10$,$500\,\AA\ \citep{2023A&A...674A...2D}. Despite their low spectral resolution, these reconstructed spectra retain sufficient information to help identify broad H, He, and Ca absorption features and infer white dwarf atmospheric compositions. Machine-learning classifications have recently produced catalogs containing on the order of 10$^5$ white dwarfs, including large numbers of candidate DZ systems \citep{2023MNRAS.521..760V, 2024A&A...682A...5V}. 

In this paper, we describe our search for massive, metal-polluted white dwarfs from candidates identified by Gaia XP spectra. This search was aimed to constrain an upper limit of progenitor host masses for surviving planetary systems. Instead, we found that Gaia XP spectra substantially increase the number of known metal-polluted white dwarfs, but masses estimated from photometric fits to the Gaia color-magnitude diagram are insufficient to identify reliable massive DZ candidates. Section~\ref{sec:sample} establishes our sample, Section~\ref{sec:observations} describes our follow-up spectroscopic observations, Section~\ref{sec:spectral types} presents the metal pollution recovery results of our study, Section~\ref{sec:masses} addresses the mass estimates of our candidates, Section~\ref{sec:discussion} discusses the new massive metal-polluted white dwarfs from our study, and Section~\ref{sec:conclusion} concludes on the success of our search for massive remnant planetary system hosts.

\vspace{-4mm}
\section{Sample Selection} \label{sec:sample}

\begin{figure*}[t]
  \centering
  {\includegraphics[width=0.9\textwidth]{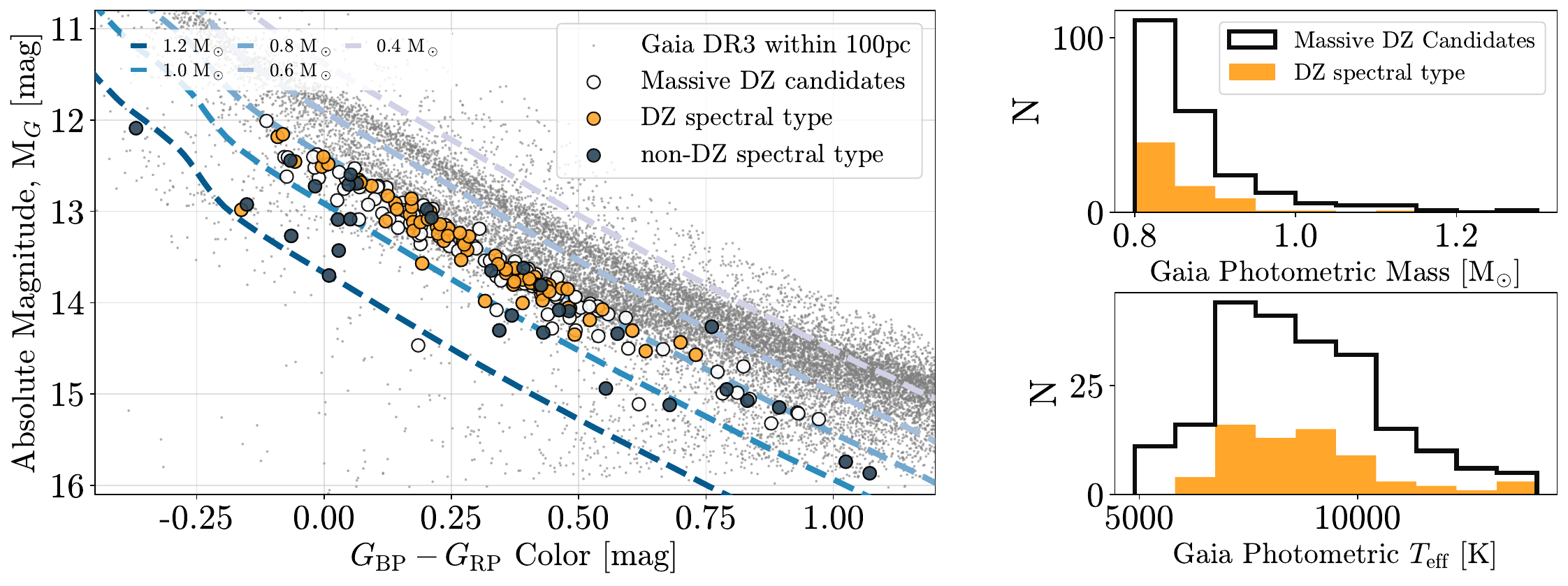}}
  \caption{\textit{Left}: Color-magnitude diagram of our sample of massive metal-polluted candidates (circles) plotted over Gaia white dwarfs within 100\,pc (gray points). The filled circles show the spectroscopically followed-up white dwarfs, with the 66 confirmed DZ white dwarfs in orange and 31 non-DZ white dwarfs in dark blue. The magnitudes of our sample are de-reddened using the coefficients adopted by \citet{2021MNRAS.508.3877G}, based on the interstellar dust extinction law of \citet{2011ApJ...737..103S}. White dwarfs of a given mass are expected to cool from the top left to the bottom right along theoretical He-atmosphere cooling tracks shown here (blue gradient lines) for masses 0.4 M$_{\odot}$ to 1.2 M$_{\odot}$. \textit{Right}: Gaia photometric mass and effective temperature parameters for our sample derived from He-dominated atmospheric models by \citet{2021MNRAS.508.3877G} (black histogram). The orange histogram shows the confirmed DZ white dwarfs. 
  } \label{fig:cmd_sample}
\end{figure*}

We selected metal-polluted candidates from the Gaia XP white dwarf spectral classification catalog of \citet{2024A&A...682A...5V}. 
Gaia XP spectra were generated for roughly 102${,}$000 high-confidence white dwarf sources across the color-magnitude diagram \citep{2023A&A...674A..33G}. 
The \citet{2024A&A...682A...5V} catalog applies a machine-learning classifier described in \citet{2023MNRAS.521..760V}, which was trained on archival SDSS spectroscopy of white dwarfs \citep{2009ApJS..182..543A, 2013ApJS..204....5K}, and predicts probabilities for typical white dwarf spectral types (DA, DB, DC, DO, DQ, and DZ) from reconstructed Gaia XP spectra. For each source, the catalog reports the probability associated with each spectral class and adopts the highest-probability class as the dominant spectral type. Among the white dwarf spectral classes, cool, metal-polluted DZ white dwarfs are particularly challenging to identify in Gaia XP spectra because the defining metal absorption features (predominantly Ca~{\sc ii}~H\,\&\,K lines at rest wavelengths of 3932\AA\ and 3969\AA) are only weakly resolved at XP spectral resolution \citep{2023MNRAS.521..760V, 2024A&A...682A...5V}. Recovery tests using archival SDSS spectra showed that the classifier described in \citet{2023MNRAS.521..760V} achieves high recall ($\sim85$\%) but modest precision ($\sim60$\%) for metal-polluted white dwarfs.

From the 102${,}$000 high confidence white dwarfs in \citet{2024A&A...682A...5V}, we select the objects whose dominant classification is DZ with probabilities assigned by the classifier to the DZ spectral class, $ P_{\rm DZ}$, ranging from 0 to 1. It is the first large all-sky sample of DZ candidates identified from Gaia XP spectroscopy alone\footnote{DZ candidates have been identified from XP spectra using unsupervised machine-learning by \citet{2024ApJ...970..181K}.}, providing a substantial increase in the number of candidate metal-polluted white dwarfs available for follow-up observations. The catalog contains 1822 metal-polluted white dwarf candidates (150 of which are included in PEWDD; \citealt{2024A&A...691A.352W}), and 1232 high confidence metal-polluted candidates with $P_{\rm DZ} >$ 0.65. Candidates with $P_{\rm DZ} <$ 0.65 have an assigned spectral type class of ``DZ:''.

In order to select massive candidates, we apply a cut on white dwarf mass from available Gaia photometry estimates, which rely on atmospheric cooling model fits. 
Above \teff\,$\gtrsim$ 5000\,K, DZ white dwarfs are believed to have He-dominated atmospheres, as their optical spectra are characterized primarily by Ca absorption lines and lack detectable Balmer absorption \citep{2007ApJ...663.1291D}, so the masses of DZ candidates are estimated assuming He-dominated atmospheric models with Gaia photometry \citep{2021MNRAS.508.3877G}. We only include candidates with masses greater than 0.8\,M$_{\odot}$, corresponding to the high-mass tail of the field white dwarf mass distribution (e.g. \citealt{2016MNRAS.461.2100T}), and which result from progenitors with masses greater than 3.5\,M$_{\odot}$ on the main sequence based on initial-final-mass-relations \citep{2022ApJ...934..148H, 2024MNRAS.527.3602C}. Of the 1822 metal-polluted white dwarf candidates, 216 objects have $M_{\rm He} >$ 0.8\,\msun\, determined by \citet{2021MNRAS.508.3877G}. 

Our selection is only sensitive to He-dominated, metal-polluted white dwarfs (DZ) and excludes H-dominated, metal-polluted white dwarfs (DAZ), which would generally be classified as DA white dwarfs in Gaia XP spectra because the Balmer absorption lines dominate over the weak metal lines. Metal lines in DAZ white dwarfs are typically narrow and difficult to resolve in low-resolution spectroscopy. Below \teff\,$\lesssim$5000\,K, DZ white dwarfs can have either  H- or He-dominated atmospheres, as the H lines are no longer visible \citep{2019ApJ...878...63B, 2024Ap&SS.369...43B}. 

Our final sample contains 216 massive, metal-polluted white dwarf candidates across the full $P_{\rm DZ}$ confidence range, including 16 candidates with $M_{\rm He} >$ 1.0\,M$_{\odot}$. 15 of the 216 massive, metal-polluted candidates are already included in PEWDD \citep{2024A&A...691A.352W}.
Figure~\ref{fig:cmd_sample} shows our candidates on a color-magnitude diagram along with the Gaia photometric mass distribution and Gaia photometric $T_{\rm eff}$ distribution, ranging from 5000 to 15${,}$000\,K. 

\section{Observations}\label{sec:observations}

\begin{figure}[t]
  \centering
  {\includegraphics[width=0.45\textwidth]{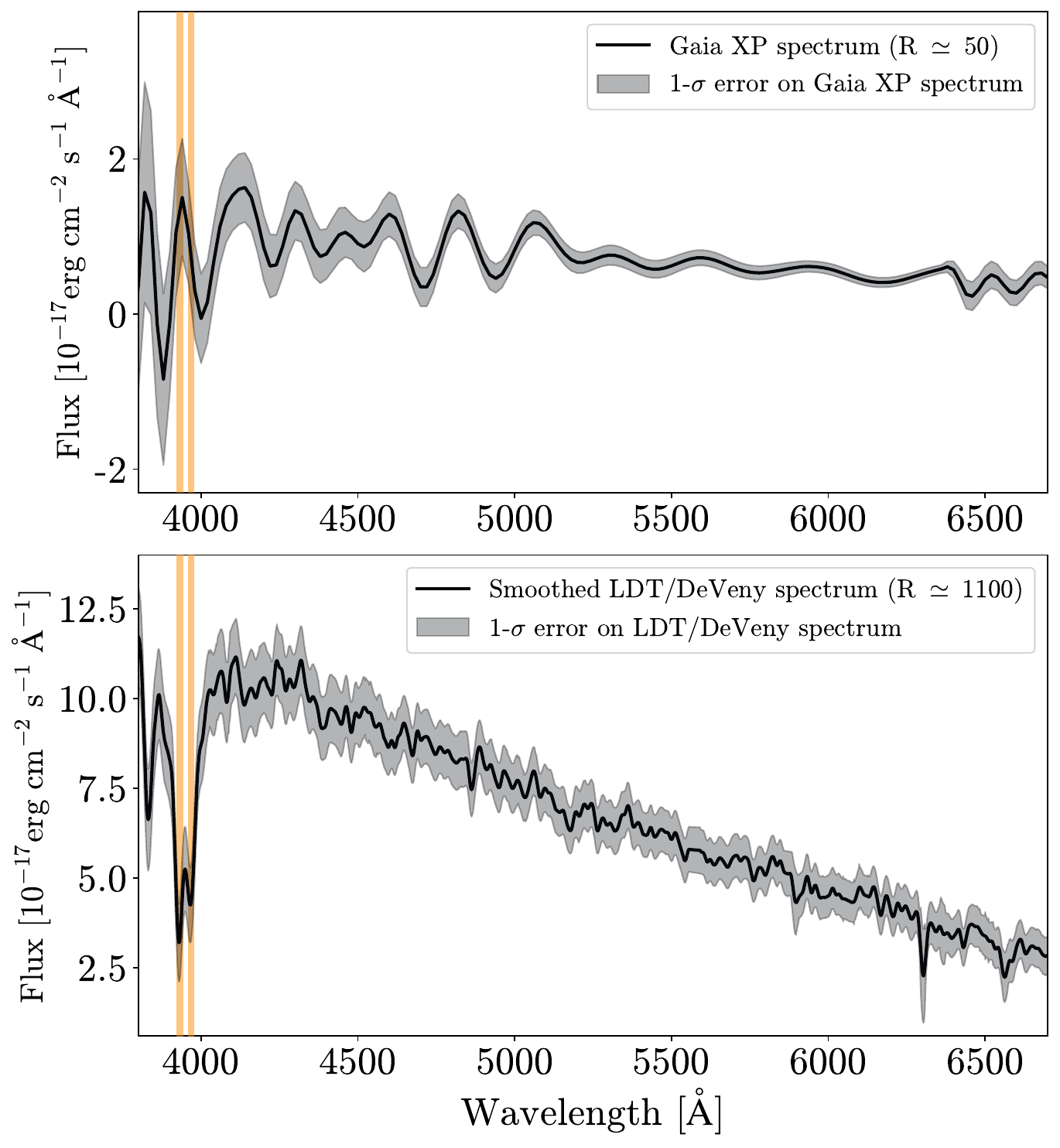}}
  \caption{Gaia XP spectrum (R$\simeq$50) of WD\,J195730.00+344812.86 (top) compared to a higher-resolution LDT/DeVeny spectrum (R$\simeq$1100) of the same target (bottom). The XP spectrum of this target gives the most likelihood to a DZ spectral type based on the methods described in \citet{2024A&A...682A...5V}, and has $P_{\rm DZ}$ = 0.55. This $G$ = 19.2 mag white dwarf is the most massive confirmed DZA of our sample (0.871~$\pm$~0.034\,M$_{\odot}$ from the fits described in Section~\ref{sec:masses}). The orange lines highlight the Ca~{\sc ii}~H\,\&\,K absorption lines, which we use to confirm the presence of photospheric metals. For each spectrum, we show a 1-$\sigma$ uncertainty contour. The XP spectrum is reconstructed from the \texttt{GaiaXPy}$^1$ package \citep{2023A&A...674A..33G}. } \label{fig:xpspec}
\end{figure}

We obtained ground-based optical spectroscopy for 40 out of the 216 in the sample, which combined with archival observations gave rise to spectroscopic observations for 97 of the 216 massive DZ candidates.  Our spectroscopic sample is not yet complete, but was determined by the availability of published and archival spectra, and observability constraints for follow-up spectroscopic campaigns. 
All 97 objects in our spectroscopically confirmed subset are shown as filled circles in Figure~\ref{fig:cmd_sample}. Follow-up observations preferentially selected targets accessible from northern facilities and brighter than $G \simeq$\,19.5, where sufficient signal-to-noise (S/N) can be obtained with reasonable exposure times ($t \approx 600$\,s at a 4-m class facility). Consequently, the spectroscopic sample may under-represent the faintest and most distant objects of our sample. We aimed to spectroscopically follow-up white dwarfs across the full cooling range of our massive metal-polluted candidate sample (along the cooling tracks shown in Figure~\ref{fig:cmd_sample}). Our spectroscopic sample of 97 objects spans the same range of  $P_{\rm DZ}$, white dwarf mass, and effective temperature as the full candidate sample of 216 objects.

Low-resolution spectroscopy is sufficient to identify absorption lines and infer spectral types based on the dominating element present in the atmosphere  \citep{1983ApJ...269..253S}. For the purposes of this work, spectra with a resolution of order $R\approx1000$ are sufficient to identify the Ca~{\sc ii}~H\,\&\,K lines that are characteristic of DZ white dwarfs. Figure~\ref{fig:xpspec} compares a Gaia XP spectrum ($R\simeq$50) to a higher-resolution Lowell Discovery Telescope DeVeny spectrum ($R\simeq$1100) of the same target, illustrating how the broad spectral features identified in the XP data correspond to the resolved Ca absorption lines in follow-up spectroscopy. The XP spectrum is reconstructed from basis functions using the \texttt{GaiaXPy}\footnote{\url{https://gaia-dpci.github.io/GaiaXPy-website}} package \citep{2023A&A...674A..33G}. 

We recovered previously published spectra for 39 candidates from our sample with the Montreal White Dwarf Database (MWDD; \citealt{2017ASPC..509....3D}) as published in the literature by \citet{2001ApJ...558..761R, 2019ApJ...885...74C, 2023MNRAS.519.4529C, 2023MNRAS.520.6135H, 2025ApJ...979..157K, 2025ApJ...990...62K}, as well as archival SDSS \citep{2013ApJS..204....5K}. 
An additional 18 spectra were obtained from DESI DR1 \citep{2026AJ....171..285D}. 
We exclude spectra that did not cover the wavelength region containing Ca~{\sc ii}~H\,\&\,K for the purpose of our analysis. We also exclude spectra with S/N $< 2$ measured in each spectrum in a 200-\AA\ wide window centered near 4600\,\AA. 

We obtained new follow-up optical spectroscopy for 40 candidates using the LDT/DeVeny, SOAR/Goodman, and MagE spectrographs. The majority of follow-up observations were obtained with DeVeny on the 4.3-m Lowell Discovery Telescope (LDT; \citealt{2014SPIE.9147E..2NB}), covering 3600$-$7400\,\AA\, using the 300 g/mm grating with central wavelength 5200\AA\ that provides spectral resolution of R$\simeq$1500 using the $1\farcs$ slit, and R$\simeq$500 using the $3\farcs$ slit (DV2 setup). Additional observations were acquired with Goodman on the 4.1-m Southern Astrophysical Research Telescope (SOAR; \citealt{2004SPIE.5492..331C}), covering 3800$-$7000\,\AA\, with the 400 line/mm grating using a $1\farcs$ slit at resolution R$\simeq$1000. One target was also observed with the Magellan Echellette (MagE) spectrograph on the Magellan Baade telescope (R$\simeq$4000).
The LDT/DeVeny, SOAR/Goodman, and MagE data were reduced and flux calibrated with the \texttt{PypeIt} pipeline \citep{2020JOSS....5.2308P, 2020zndo...3743493P}. Due to electronic interference on some DeVeny exposures, we also use the DeVeny Pickup Noise Scrubber\footnote{\url{https://lowellobservatory.github.io/LDTObserverTools/}} to mitigate interference patterns in the CCD bias. 
Our observation details, including dates and setups used are described in Table~\ref{tab:obs}. 

\begin{figure}[t]
  \centering
  {\includegraphics[width=.44\textwidth]{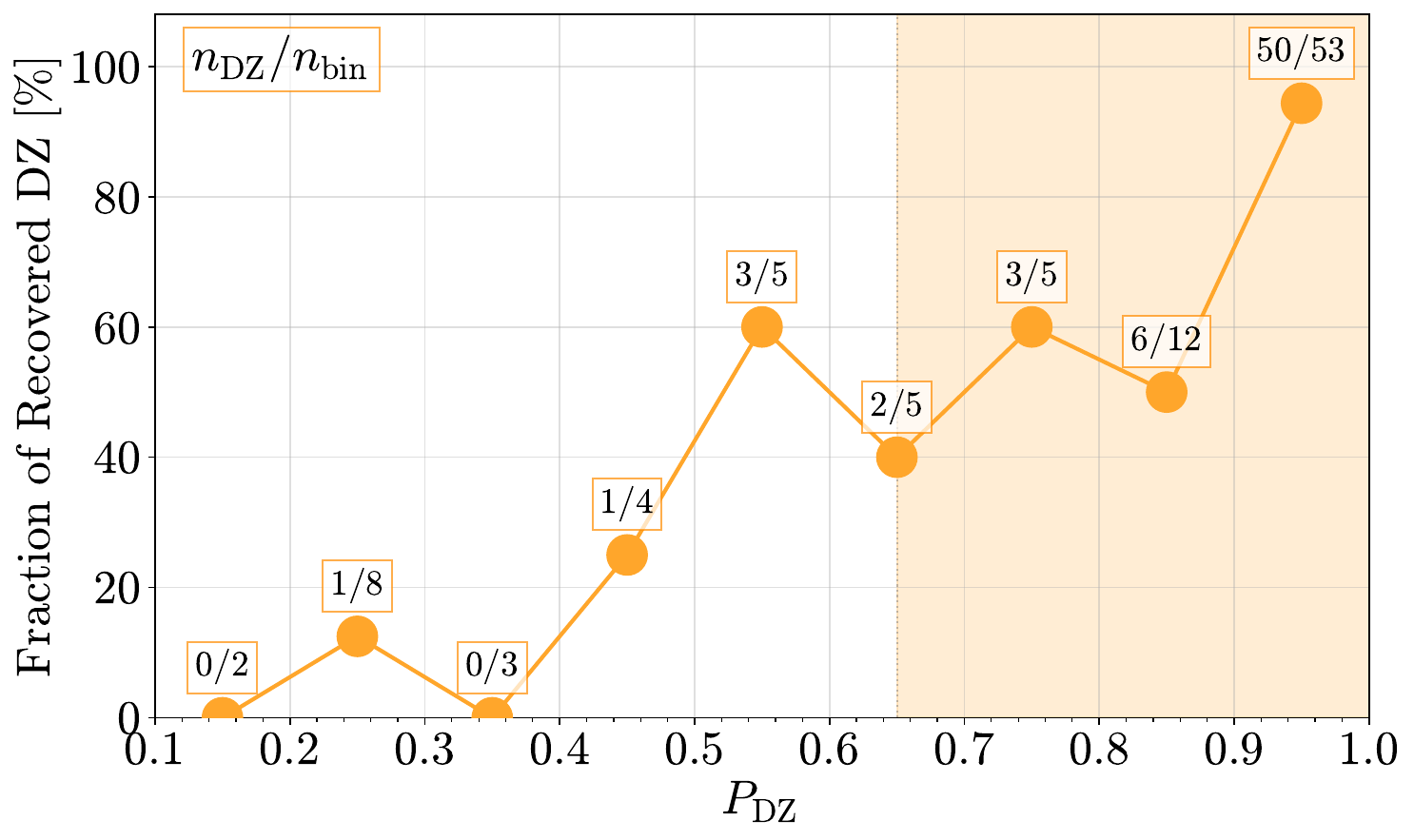}}
  \caption{Fraction of recovered metal-polluted (DZ) white dwarfs as a function of the classifier probability, $P_{\rm DZ}$ from \citet{2024A&A...682A...5V}. The markers are centered on each bin. The boxed numbers above each marker indicate the size of that bin. The shaded region reflects the high confidence DZ candidates, for which we recover 82~$\pm$~5\%, or 59 out of the 72 targets with $P_{\rm DZ}> 0.65$. Recovery is most effective at higher probabilities, reaching 94~$\pm$~3\% for $P_{\rm DZ}> 0.9$.} \label{fig:pdz_fraction}
\end{figure}

\section{Spectral Types of Massive DZ Candidates} \label{sec:spectral types}

We assign spectral classifications for each object of our spectroscopic subset of 97 massive metal-polluted white dwarf candidates through visual inspection of the reduced spectra, based primarily on the presence of Ca~{\sc ii}~H\,\&\,K absorption at rest wavelengths of 3969\,\AA\ and 3934\,\AA, respectively. The observed metals are predicted to be photospheric accretion rather than interstellar contamination based on the broadness of the features from the strong surface gravity of our sample (e.g., \citealt{2007ApJ...663.1291D}). 
We detect photospheric Ca~{\sc ii}~H\,\&\,K in 66 white dwarfs of our sample, which we present in Table~\ref{tab:dzs}. Specifically, 58 only show photospheric metal lines and are classified as DZ, six also show marginal absorption at the H Balmer lines and are classified as DZA, while one additional white dwarf shows strong H absorption and is classified as DAZ. Finally, one white dwarf, WD\,J010728.61+265021.28, shows metal absorption with Zeeman splitting, characteristic of strong magnetism, and is therefore classified as DZH.
For the purpose of our analysis, we refer to any white dwarf with detectable metal pollution as DZ. The precise spectral type of each object is listed in Table~\ref{tab:dzs}. 

We recover 66 DZ white dwarfs out of the 97 spectroscopic candidates, or 68~$\pm$~5\%, with uncertainties derived from binomial statistics. Only considering high-confidence candidates, with $P_{\rm DZ} > 0.65$, we recover 82~$\pm$~5\% DZ white dwarfs (59/72). Figure~\ref{fig:pdz_fraction} shows the fraction of recovered DZ white dwarfs as a function of $P_{\rm DZ}$, in which it can be seen that the confirmed metal-polluted white dwarfs span the full range of $ P_{\rm DZ}$. The recovery fraction increases toward higher classifier probabilities, reaching 94~$\pm$~3\% recovery (50/53) for candidates with $P_{\rm DZ}>0.9$. We note that uncertainties on fractional recovery do not include possible systematic effects from spectroscopic completeness or classification ambiguity near the detection threshold.

\begin{figure}[t]
  \centering
  {\includegraphics[width=0.451\textwidth]{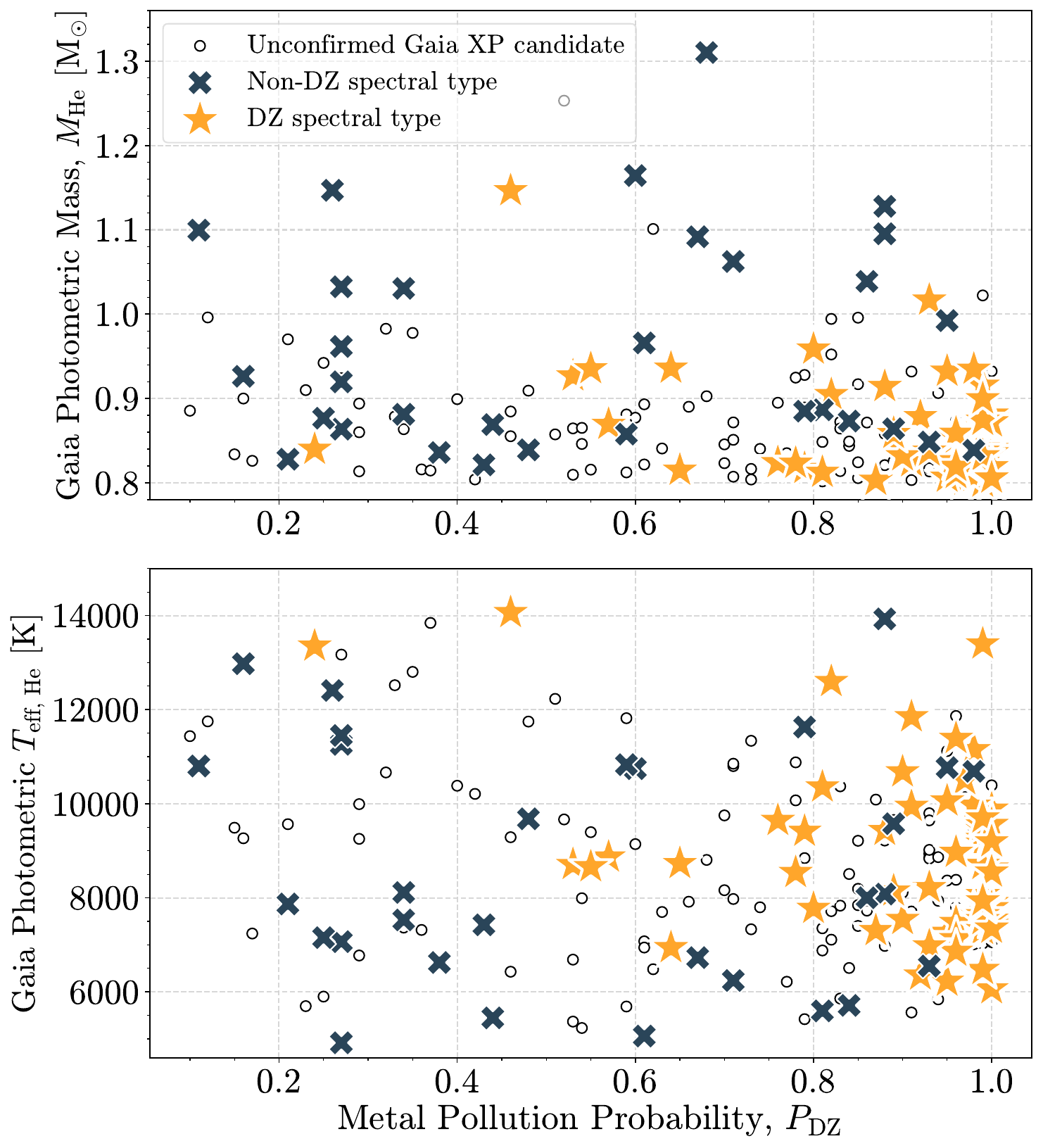}}
  \caption{\textit{Top:} Gaia photometric white dwarf mass from He-atmospheric models, $M_{\rm He}$, as a function of the classifier probability of being metal polluted ($P_{\rm DZ}$) \citep{2021MNRAS.508.3877G, 2024A&A...682A...5V}. The orange stars represent the 66 spectroscopically confirmed DZ white dwarfs. The black cross markers represent the 31 white dwarfs which do not show Ca~{\sc ii}~H\,\&\,K lines (non-DZ white dwarfs). The majority of non-DZ are continuum-dominated white dwarfs (23/31), see Table~\ref{tab:non-dzs}. The background open circles show the remaining unconfirmed Gaia XP massive metal-polluted candidates.
  \textit{Bottom:} Similar to the top panel, we now show the Gaia photometric white dwarf effective temperature, $T_{\rm eff,\,He}$ from He-atmospheric models as a function of $P_{\rm DZ}$.} \label{fig:mass_pdz}
\end{figure}

In Figure~\ref{fig:mass_pdz}, we compare the Gaia photometric mass, $M_{\rm He}$, and Gaia photometric effective temperature $T_{\rm eff}$ for confirmed DZ white dwarfs and non-DZ white dwarfs, as derived from Gaia photometry under the assumption of a He-dominated atmosphere. There are no significant correlations between either mass or temperature and recovery rate. Still, only 2/13 white dwarfs with Gaia photometric mass $M_{\rm He}>$ 1.0\,\msun\, are recovered as metal-polluted (15~$\pm$~10\%), compared to 76~$\pm$~5\% (64/84) in the 0.8-1.0\,\msun\ range. As $T_{\rm eff,\, He}$ gets cooler, the recovery fraction also appears to decrease. None of the ten candidates with $P_{\rm DZ}<0.5$ and $T_{\rm eff} < 10{,}000$ show metal pollution.

\begin{longrotatetable}
\begin{deluxetable*}{lccccccccr}
\tablecaption{Confirmed Metal-Polluted White Dwarfs (DZ)\label{tab:dzs}}
\tablecolumns{10}
\tabletypesize{\footnotesize}

\tablehead{
    \colhead{Target Name} &
    \colhead{Gaia mass, $M_{\rm He}$} &
    \colhead{$P_{\rm DZ}$} &
    \colhead{Mass, $M_{\rm DZ}$} &
    \colhead{$T_{\rm eff,\, DZ}$} &
    \colhead{$\log(g)_{\rm DZ}$} &
    \colhead{$\log\rm(Ca/He)$} &
    \colhead{$\log\rm(H/He)$} & 
    \colhead{Spec Source} &
    \colhead{Spectral Type}
}
\startdata
WD\,J002617.27+001543.38 & 1.016 (0.143) & 0.93 & 0.755 (0.057) & 6550 (70) & 8.28 (0.06) & -9.16 & -5.0 & This Work & DZ\\
WD\,J003908.19+562002.49 & 0.828 (0.082) & 0.99 & 0.720 (0.044) & 7010 (190) & 8.23 (0.05) & -9.72 & -5.0 & This Work & DZ\\
WD\,J004809.33-012438.60 & 0.840 (0.028) & 1.00 & 0.729 (0.007) & 8770 (40) & 8.24 (0.01) & -8.30 & -4.0 & DESI & DZA\\
WD\,J010728.61+265021.28 & 0.820 (0.096) & 1.00 & 0.688 (0.026) & 5450 (50) & 8.19 (0.03) & -8.86 & -3.0 & SDSS & DZH\\
WD\,J012444.74-222906.88 & 0.817 (0.024) & 0.98 &  &  &  &  &  & \citet{2023MNRAS.519.4529C} & DZ\\
WD\,J032200.53+801349.44 & 0.807 (0.074) & 0.95 & 0.688 (0.018) & 5540 (30) & 8.19 (0.02) & -8.65 & -5.0 & \citet{2025ApJ...979..157K} & DZ\\
WD\,J032833.52-121945.27 & 0.876 (0.039) & 1.00 & 0.672 (0.011) & 7380 (70) & 8.16 (0.01) & -9.52 & -4.0 & This Work & DZ\\
WD\,J035053.31+660433.38 & 0.936 (0.094) & 0.64 & 0.809 (0.040) & 6180 (120) & 8.36 (0.04) & -9.51 & -4.0 & This Work & DZ\\
WD\,J035345.55+401426.17 & 0.817 (0.094) & 0.79 & 0.646 (0.040) & 8050 (190) & 8.11 (0.04) & -8.89 & -5.0 & This Work & DZ\\
WD\,J055742.01+175256.25 & 0.803 (0.163) & 0.97 & 0.604 (0.046) & 8460 (150) & 8.05 (0.05) & -8.18 & -5.0 & This Work & DZ\\
WD\,J064735.80+705803.59 & 0.803 (0.053) & 1.00 & 0.648 (0.020) & 6510 (80) & 8.12 (0.02) & -9.15 & -5.0 & DESI & DZ\\
WD\,J073418.29+672801.13 & 0.808 (0.137) & 0.99 & 0.625 (0.055) & 8090 (200) & 8.08 (0.06) & -9.64 & -5.0 & DESI & DZ\\
WD\,J075307.36+661354.18 & 0.866 (0.170) & 0.99 & 0.604 (0.068) & 7630 (120) & 8.05 (0.08) & -8.89 & -4.0 & SDSS & DZ\\
WD\,J080341.32+651848.46 & 0.932 (0.162) & 0.95 & 0.737 (0.062) & 8170 (110) & 8.25 (0.07) & -8.04 & -4.0 & DESI & DZA\\
WD\,J083858.55+232252.98 & 0.819 (0.085) & 0.99 & 0.694 (0.022) & 5920 (40) & 8.19 (0.02) & -9.54 & -3.0 & SDSS & DZ\\
WD\,J084849.43+354858.07 & 0.869 (0.055) & 0.57 & 0.681 (0.016) & 7860 (50) & 8.17 (0.02) & -10.39 & -4.0 & SDSS & DZ\\
WD\,J084911.87+403649.75 & 0.812 (0.064) & 0.81 & 0.648 (0.021) & 9180 (70) & 8.11 (0.02) & -10.29 & -5.0 & SDSS & DZ\\
WD\,J085749.95+263026.96 & 0.833 (0.153) & 1.00 & 0.475 (0.047) & 6310 (80) & 7.83 (0.06) & -10.72 & -5.0 & SDSS & DZ\\
WD\,J091714.53+263015.31 & 0.904 (0.106) & 0.82 & 0.774 (0.035) & 11040 (130) & 8.27 (0.04) & -8.98 & -4.0 & SDSS & DAZ\\
WD\,J093412.31+563232.54 & 0.853 (0.103) & 0.96 & 0.605 (0.028) & 6250 (60) & 8.05 (0.03) & -9.99 & -3.0 & SDSS & DZ\\
WD\,J095119.87+403322.57 & 0.808 (0.030) & 0.99 & 0.686 (0.008) & 8210 (50) & 8.17 (0.01) & -10.03 & -4.0 & SDSS & DZ\\
WD\,J111215.06+070052.57 & 0.851 (0.191) & 0.99 & 0.500 (0.058) & 6980 (60) & 7.87 (0.07) & -9.06 & -4.0 & SDSS & DZ\\
WD\,J112144.68+390704.68 & 0.855 (0.196) & 0.99 & 0.669 (0.064) & 10670 (180) & 8.14 (0.07) & -7.79 & -5.0 & SDSS & DZA\\
WD\,J112206.33+842647.72 & 0.836 (0.073) & 1.00 & 0.766 (0.061) & 6980 (340) & 8.30 (0.07) & -9.00 & -3.0 & This Work & DZ\\
WD\,J112807.93+530625.77 & 0.879 (0.068) & 0.92 & 0.674 (0.019) & 5500 (40) & 8.17 (0.02) & -10.38 &  & This Work & DZ\\
WD\,J115926.98+290105.85 & 0.824 (0.140) & 0.76 & 0.793 (0.052) & 9400 (110) & 8.34 (0.06) & -10.34 & -6.0 & SDSS & DZ\\
WD\,J120556.55+724808.13 & 0.809 (0.132) & 0.99 & 0.679 (0.035) & 7090 (130) & 8.17 (0.04) & -8.88 & -5.0 & DESI & DZ\\
WD\,J120840.87+190710.23 & 0.823 (0.144) & 0.78 & 0.608 (0.039) & 7440 (80) & 8.05 (0.04) & -10.43 & -5.0 & SDSS & DZ\\
WD\,J122254.10+233229.48 & 0.916 (0.114) & 0.99 & 0.761 (0.036) & 8570 (120) & 8.29 (0.04) & -9.56 & -4.0 & SDSS & DZA\\
WD\,J122314.50+800225.01 & 0.807 (0.105) & 0.95 & 0.718 (0.028) & 6630 (100) & 8.23 (0.03) & -9.15 & -4.0 & This Work & DZ\\
WD\,J130413.85+045457.71 & 0.935 (0.166) & 0.98 & 0.660 (0.058) & 6210 (50) & 8.14 (0.06) & -9.40 & -5.0 & DESI & DZ\\
WD\,J132147.14+161446.55 & 0.826 (0.087) & 0.96 & 0.680 (0.027) & 6710 (70) & 8.17 (0.03) & -9.62 & -3.0 & SDSS & DZ\\
WD\,J135217.75+552338.58 & 0.835 (0.181) & 1.00 & 0.664 (0.037) & 6760 (80) & 8.14 (0.04) & -9.31 & -5.0 & DESI & DZ\\
WD\,J135637.80+404703.46 & 0.838 (0.085) & 0.99 & 0.645 (0.022) & 6820 (50) & 8.12 (0.02) & -9.65 & -4.0 & SDSS & DZ\\
WD\,J140207.64+251827.30 & 0.801 (0.133) & 0.99 & 0.770 (0.042) & 6940 (100) & 8.30 (0.05) & -9.71 &  & SDSS & DZ\\
WD\,J140918.90+151136.96 & 0.803 (0.081) & 0.87 & 0.579 (0.019) & 6410 (40) & 8.01 (0.02) & -10.88 & -5.0 & SDSS & DZ\\
WD\,J143641.81+295138.70 & 0.881 (0.092) & 1.00 & 0.632 (0.024) & 7630 (70) & 8.09 (0.03) & -8.80 & -5.0 & SDSS & DZ\\
WD\,J144124.36+083104.68 & 0.871 (0.121) & 0.99 & 0.547 (0.021) & 7100 (50) & 7.95 (0.03) & -10.78 & -4.0 & SDSS & DZ\\
WD\,J144804.48+104709.20 & 0.872 (0.115) & 1.00 & 0.659 (0.021) & 6280 (30) & 8.14 (0.02) & -9.24 & -5.0 & SDSS & DZ\\
WD\,J144830.62+295817.51 & 0.841 (0.122) & 0.99 & 0.699 (0.045) & 8620 (140) & 8.19 (0.05) & -9.26 & -5.0 & SDSS & DZ\\
WD\,J152224.25+103951.36 & 0.914 (0.162) & 0.88 & 0.728 (0.046) & 7920 (110) & 8.24 (0.05) & -8.80 & -5.0 & This Work & DZ\\
WD\,J152505.42+303431.61 & 0.861 (0.113) & 0.96 & 0.791 (0.028) & 6620 (60) & 8.34 (0.03) & -9.35 & -5.0 & DESI & DZ\\
WD\,J153538.31+383539.18 & 0.824 (0.151) & 0.99 & 0.564 (0.039) & 6970 (80) & 7.98 (0.05) & -8.78 & -5.0 & DESI & DZ\\
WD\,J153746.70+301728.17 & 0.808 (0.089) & 0.98 & 0.693 (0.027) & 9530 (100) & 8.18 (0.03) & -7.92 & -6.0 & This Work & DZ\\
WD\,J155850.74+091604.62 & 0.840 (0.177) & 0.24 & 0.610 (0.060) & 10600 (190) & 8.05 (0.07) & -9.89 &  & DESI & DZ\\
WD\,J160429.81+183035.58 & 0.837 (0.184) & 0.99 & 0.601 (0.048) & 6410 (70) & 8.05 (0.05) & -9.70 & -5.0 & SDSS & DZ\\
WD\,J161114.31+111708.27 & 0.825 (0.126) & 0.91 & 0.595 (0.038) & 8010 (80) & 8.03 (0.04) & -8.57 & -5.0 & SDSS & DZ\\
WD\,J161444.18+133414.68 & 0.811 (0.132) & 0.96 & 0.685 (0.033) & 7970 (80) & 8.17 (0.04) & -8.84 & -5.0 & SDSS & DZ\\
WD\,J162458.18+540429.97 & 0.819 (0.062) & 0.96 & 0.775 (0.014) & 10530 (100) & 8.31 (0.02) & -8.04 & -6.0 & This Work & DZ\\
WD\,J174213.13+155719.84 & 0.841 (0.108) & 0.90 & 0.624 (0.037) & 6490 (120) & 8.08 (0.04) & -9.91 & -5.0 & This Work & DZ\\
WD\,J175924.70+285613.99 & 0.819 (0.156) & 1.00 & 0.380 (0.056) & 6710 (270) & 7.63 (0.09) & -9.19 & -5.0 & DESI & DZ\\
WD\,J185045.15-025939.58 & 0.927 (0.107) & 0.53 & 0.670 (0.031) & 6940 (100) & 8.15 (0.03) & -8.56 & -3.0 & This Work & DZ\\
WD\,J192632.10+560039.45 & 0.825 (0.126) & 0.91 & 0.653 (0.030) & 9680 (150) & 8.12 (0.03) & -8.18 & -5.0 & This Work & DZ\\
WD\,J195707.90+334441.41 & 1.146 (0.111) & 0.46 & 0.789 (0.070) & 9160 (230) & 8.33 (0.08) & -9.15 & -4.0 & This Work & DZA\\
WD\,J195730.00+344812.86 & 0.935 (0.115) & 0.55 & 0.871 (0.034) & 8100 (210) & 8.45 (0.04) & -8.38 & -3.0 & This Work & DZA\\
WD\,J195938.61-020829.97 & 0.859 (0.080) & 0.96 & 0.692 (0.020) & 5920 (50) & 8.19 (0.02) & -9.92 &  & This Work & DZ\\
WD\,J204816.09+061613.28 & 0.838 (0.167) & 0.93 & 0.675 (0.051) & 6060 (90) & 8.16 (0.06) & -9.04 & -3.0 & DESI & DZ\\
WD\,J210119.94+812246.73 & 0.802 (0.104) & 1.00 & 0.543 (0.027) & 6840 (100) & 7.95 (0.03) & -9.04 &  & This Work & DZ\\
WD\,J211120.21-051032.16 & 0.832 (0.073) & 0.90 & 0.746 (0.026) & 9470 (110) & 8.26 (0.03) & -8.21 & -6.0 & This Work & DZ\\
WD\,J213909.61-070244.82 & 0.832 (0.097) & 0.99 & 0.669 (0.030) & 6670 (60) & 8.15 (0.03) & -9.86 & -5.0 & This Work & DZ\\
WD\,J222503.65+233855.87 & 0.806 (0.150) & 1.00 & 0.658 (0.038) & 6420 (60) & 8.14 (0.04) & -9.11 & -5.0 & SDSS & DZ\\
WD\,J223257.71+241303.62 & 0.815 (0.096) & 0.65 & 0.629 (0.024) & 7700 (80) & 8.09 (0.03) & -9.92 & -5.0 & This Work & DZ\\
WD\,J223841.05+010150.44 & 0.858 (0.171) & 0.89 & 0.654 (0.045) & 7040 (70) & 8.13 (0.05) & -9.87 & -5.0 & SDSS & DZ\\
WD\,J231546.83+052000.25 & 0.958 (0.169) & 0.80 & 0.593 (0.031) & 6160 (60) & 8.03 (0.04) & -10.82 & -6.0 & SDSS & DZ\\
WD\,J231747.18+370121.20 & 0.874 (0.115) & 0.99 & 0.711 (0.032) & 6780 (110) & 8.22 (0.03) & -8.62 & -3.0 & This Work & DZ\\
WD\,J235516.58+143136.28 & 0.899 (0.105) & 0.99 & 0.638 (0.036) & 6660 (70) & 8.10 (0.04) & -10.38 & -3.0 & SDSS & DZ\\
\enddata
\end{deluxetable*}
\end{longrotatetable}

\begin{deluxetable*}{lccccr}[htbp!]
\tablecaption{False Positive Candidates (non-DZ)\label{tab:non-dzs}}
\tablecolumns{6}
\tabletypesize{\footnotesize}

\tablehead{
    \colhead{Target Name} &
    \colhead{Gaia mass, $M_{\rm He}$} &
    \colhead{Gaia $T_{\rm eff,\, He}$} &
    \colhead{$P_{\rm DZ}$} &
    \colhead{Spec Source} &
    \colhead{Spectral Type}
}

\startdata
WD\,J021205.31+064420.27 & 1.128 (0.073) & 13930 (1590) & 0.88 & SDSS & DH \\
WD\,J021553.69-155542.84 & 0.926 (0.153) & 12980 (1740) & 0.16 & \citet{2025ApJ...979..157K} & DH \\
WD\,J030327.72+031715.93 & 0.864 (0.153) & 7070 (610) & 0.27 & \citet{2025ApJ...979..157K} & DC \\
WD\,J043019.38-294503.91 & 1.100 (0.053) & 10800 (660) & 0.11 & This Work & DH \\
WD\,J050823.28+744430.34 & 0.966 (0.058) & 5060 (150) & 0.61 & This Work & DC \\
WD\,J051753.43-434204.78 & 1.095 (0.089) & 8080 (670) & 0.88 & This Work & DC \\
WD\,J063645.64-545420.49 & 1.032 (0.104) & 11260 (1150) & 0.27 & This Work & Warm DQ \\
WD\,J071608.21-145652.81 & 1.164 (0.084) & 10750 (1100) & 0.60 & This Work & DA \\
WD\,J075421.27+284353.79 & 0.887 (0.150) & 5600 (370) & 0.81 & DESI & DC \\
WD\,J080042.49+065542.09 & 0.992 (0.043) & 10770 (430) & 0.95 & \citet{2025ApJ...979..157K} & DAH \\
WD\,J084343.92+035452.09 & 0.821 (0.186) & 7430 (720) & 0.43 & DESI & DA \\
WD\,J085357.69-244656.23 & 0.962 (0.028) & 4920 (70) & 0.27 & \citet{2001ApJ...558..761R} & DC \\
WD\,J085649.69+253441.17 & 0.839 (0.045) & 10690 (350) & 0.98 & \citet{2025ApJ...979..157K} & DAH \\
WD\,J112148.81+103934.33 & 0.858 (0.078) & 10830 (660) & 0.59 & SDSS & DH \\
WD\,J114305.14+080709.98 & 0.885 (0.204) & 11630 (1540) & 0.79 & This Work & DA \\
WD\,J114843.35-152814.51 & 0.869 (0.065) & 5450 (170) & 0.44 & This Work & DC \\
WD\,J130757.88+720836.58 & 1.092 (0.069) & 6730 (400) & 0.67 & This Work & DC \\
WD\,J131059.19-001306.88 & 0.839 (0.164) & 9680 (890) & 0.48 & DESI & DA \\
WD\,J141030.66-315723.61 & 0.864 (0.170) & 9580 (1000) & 0.89 & This Work & DC \\
WD\,J142018.78+035130.98 & 0.836 (0.103) & 6630 (360) & 0.38 & \citet{2025ApJ...979..157K} & DC \\
WD\,J150603.22-604016.49 & 1.310 (0.206) & 37130 (11470) & 0.68 & This Work \& MagE & DBA \\
WD\,J155842.94-235841.58 & 1.062 (0.035) & 6240 (160) & 0.71 & This Work & DC \\
WD\,J155934.77+731357.85 & 0.873 (0.074) & 5710 (230) & 0.84 & This Work & DC \\
WD\,J181819.84+152434.51 & 1.039 (0.036) & 8000 (230) & 0.86 & This Work & DC \\
WD\,J191649.94+640954.13 & 0.848 (0.141) & 6550 (500) & 0.93 & DESI & DC \\
WD\,J195919.73+583634.93 & 0.920 (0.105) & 11450 (980) & 0.27 & This Work & DH \\
WD\,J202617.90-002149.25 & 1.147 (0.100) & 12410 (1540) & 0.26 & This Work & DH \\
WD\,J205219.66-084534.72 & 1.031 (0.098) & 7530 (480) & 0.34 & DESI & DC \\
WD\,J215352.28+541408.86 & 0.881 (0.123) & 8110 (580) & 0.34 & This Work & DC \\
WD\,J224511.73+684122.76 & 0.877 (0.117) & 7150 (530) & 0.25 & This Work & DC \\
WD\,J224923.35+254955.75 & 0.828 (0.202) & 7870 (860) & 0.21 & DESI & DC \\
\enddata
\end{deluxetable*}

\vspace{-50pt}

\subsection{False Positives}\label{sec:false-pos}

Of the 97 white dwarfs in our spectroscopic sample of massive DZ candidates, 31 are confirmed to be non-DZ, and are listed in Table~\ref{tab:non-dzs}. 
These objects are predominantly continuum-dominated white dwarfs, with 23 showing no detectable spectral features at our sensitivity. Cool, continuum-dominated white dwarfs are usually classified as DC white dwarfs. Above \teff\,$>$10${,}$000\,K, H, He, or C features are expected to be visible. The warm continuum-dominated white dwarfs are likely to be strongly magnetic (field strength $B$ $\geq$ 50 MG), where non-linear Zeeman effects can severely distort or suppress spectral features, preventing reliable line identification (e.g. \citealt{2015SSRv..191..111F}). We classify these strongly-magnetic white dwarfs as DH, which may have either H- or He-dominated atmospheres. Of these 23 continuum-dominated objects, we classify 17 as DC white dwarfs and six as strongly magnetic DH white dwarfs. In these cases, the non-linear Zeeman splitting and line blending prevent reliable magnetic field estimates using standard fitting tools like the \texttt{Zeeman splitting tool}\footnote{\url{https://github.com/WD-planets/Zeeman\_splitting}} \citep{2014ApJS..212...26S, 2023MNRAS.524.4867I}. 
We also recover two strongly magnetic H-dominated systems, classified as DAH, and distinguished by magnetically-distorted Balmer line profiles. These two DAH white dwarfs, WD\,J0800+0655 and WD\,J0856+2534, have field strength $B = $317.0\,MG and $B =$85\,MG, respectively \citep{2025ApJ...990...25M}. Overall, at least eight false positive white dwarfs from our sample show strong magnetism.  
We identify a further four targets with non-Zeeman split Balmer absorption as H-dominated DA white dwarfs.

\begin{figure}[t]
  \centering
  {\includegraphics[width=0.48\textwidth]{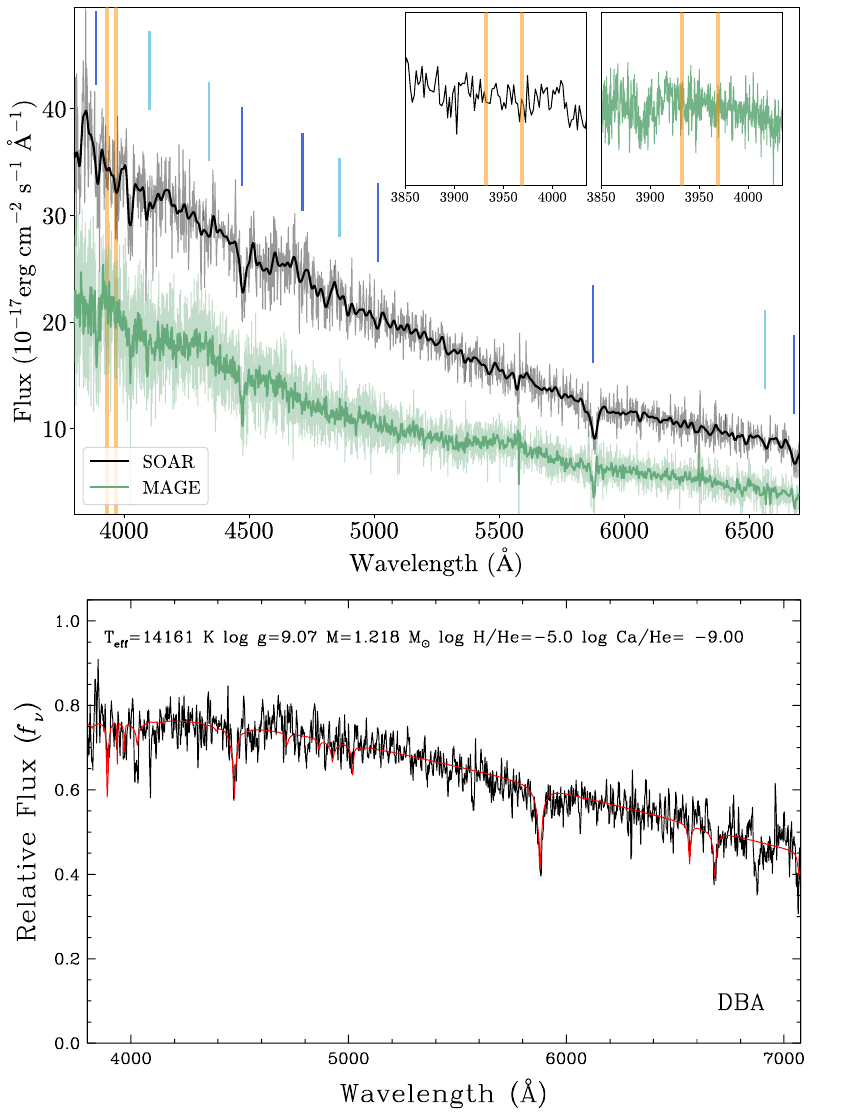}}
  \caption{\textit{Top:} Optical spectroscopy of WD\,J150603.22$-$604016.49 from SOAR/Goodman (black spectrum, S/N = 9.9) and Magellan/MagE (blue spectrum, S/N = 6.2). WD\,J150603.22$-$604016.49 is the most massive metal-polluted candidate from our sample with a Gaia photometric mass 1.31~$\pm$~0.21\,M$_{\odot}$. The insets zoom in on the Ca~{\sc ii}~H\,\&\,K region. While we detect He lines (dark blue) along with weak H Balmer lines (light blue), we do not confidently detect metals in this object, and classify this white dwarf as a DBA. Higher resolution spectroscopy is needed to more confidently rule out the presence of metals. \textit{Bottom:} Spectroscopic fit with DBA model.} \label{fig:WDJ1506}
\end{figure}

We also identify two outliers. The first, WD\,J063645.64$-$545420.49 is a metal-polluted candidate with Gaia photometric mass 1.03~$\pm$~0.10\,M$_{\odot}$. Follow-up spectroscopy with SOAR/Goodman reveals absorption features from atomic C, consistent with a C-dominated atmosphere. This target is reminiscent of warm DQ white dwarfs described in \citet{2024ApJ...974...12J}, with an unusually C-rich or C-dominated atmosphere and high transverse velocity (here, $v_t$ = 73.6~$\pm$~2.6 km\,s$^{-1}$). The best-fitting DQ model returns log~C/H~=~3.00, $T_{\rm eff} =$ 11,770~$\pm$~300\,K and $M_{\rm WD}=$1.088~$\pm$~0.028\,\msun, based on the method described in \citet{2026arXiv260625377K} and photometry from the Dark Energy Survey (DES; \citealt{2021ApJS..254...24S}). The full fit is included in Figure~\ref{fig:dqfit}. 
The second outlier, WD\,J150603.22$-$604016.49 is a metal-polluted candidate with Gaia photometric mass 1.31~$\pm$~0.21\,M$_{\odot}$. It stood out as the most massive candidate in our sample. Follow-up spectroscopy with SOAR/Goodman (S/N = 9.9; black spectrum in top panel of Fig.~\ref{fig:WDJ1506}) showed absorption lines from He and H, and possibly hints of Ca~{\sc ii}. We obtained further follow-up optical spectroscopic observations of this target with Magellan/MagE, in an effort to confirm the presence of Ca~{\sc ii} absorption, but the acquired spectrum (S/N = 6.2; blue spectrum in top panel of Fig.~\ref{fig:WDJ1506}) remains ambiguous. For the purpose of this study, we classify this white dwarf as a DBA. Higher resolution spectroscopy is required to confirm the presence of photospheric metals. This object is not included in publicly available photometric surveys, but synthetic photometric fits from XP spectra from \citet{2024A&A...682A...5V} return a mass of 0.731~$\pm$~0.152\,M$_{\odot}$. The bottom panel of Figure~\ref{fig:WDJ1506} shows our spectroscopic fit to the SOAR/Goodman spectrum with DBA models, which returns a mass of 1.218~$\pm$~0.021\,M$_{\odot}$. 

Overall, the 31 false positives of our sample include 17 continuum-dominated white dwarfs (DC), eight strongly magnetic white dwarfs (DH and DAH), four non-magnetic H-dominated white dwarfs (DA), one C-dominated white dwarf (warm DQ), and one mixed atmosphere white dwarf with H and He (DBA). All new spectra of the non-DZ white dwarfs are presented in Figures~\ref{fig:WDJ1506}, \ref{fig:dqfit}, \ref{fig:tab:nonDZs-spec}, \ref{fig:tab:nonDZs-spec-2}, and \ref{fig:tab:nonDZs-spec-3}.

\figsetgrpstart
\figsetgrpnum{6.54}
\figsetgrptitle{WDJ195730.00+344812.86}
\figsetplot{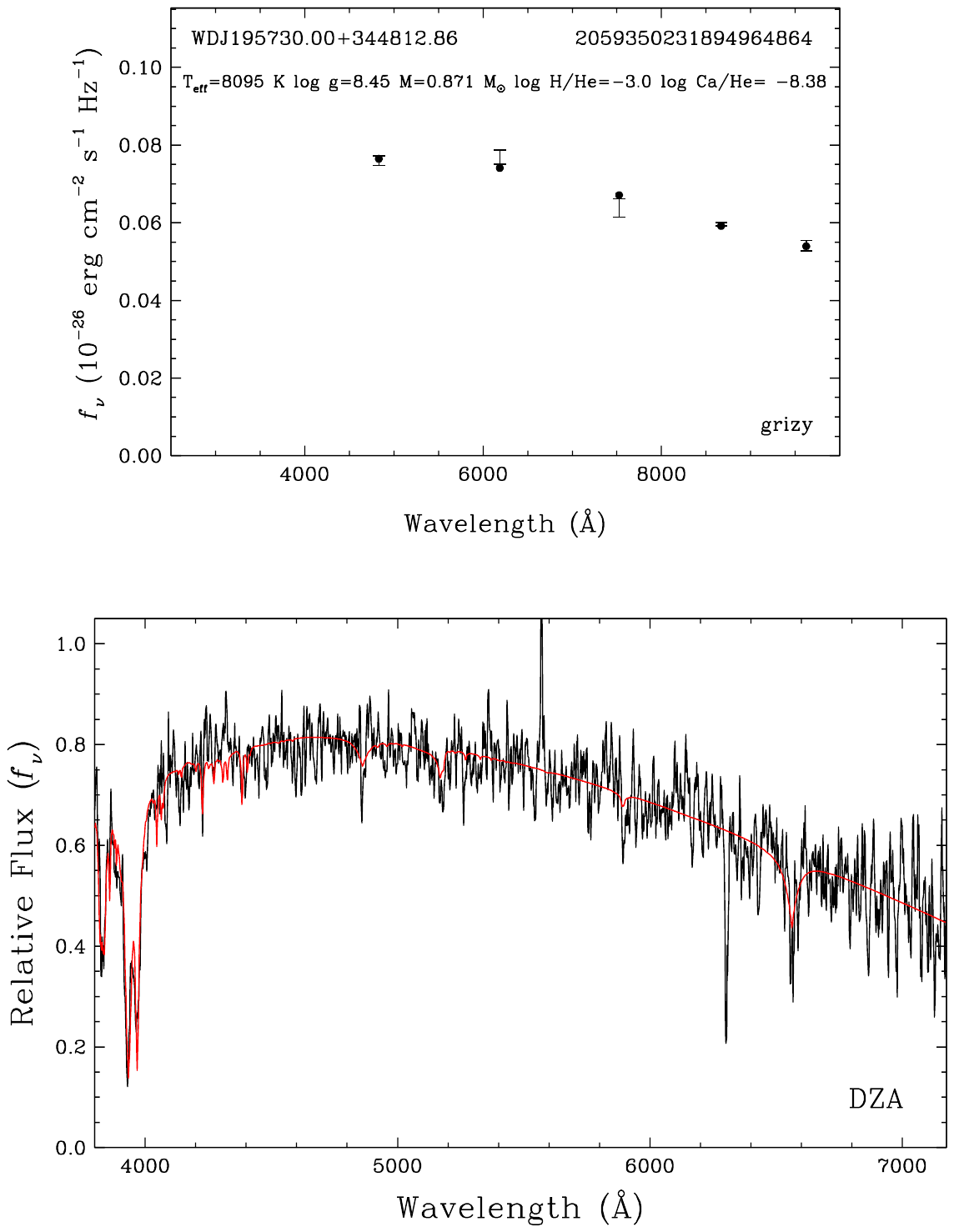}
\figsetgrpnote{Photometric spectral energy distribution and optical spectrum with best-fit atmospheric model for WDJ195730.00+344812.86.}
\figsetgrpend

\figsetend

\begin{figure}[t]
  \centering
  {\includegraphics[width=0.48\textwidth]{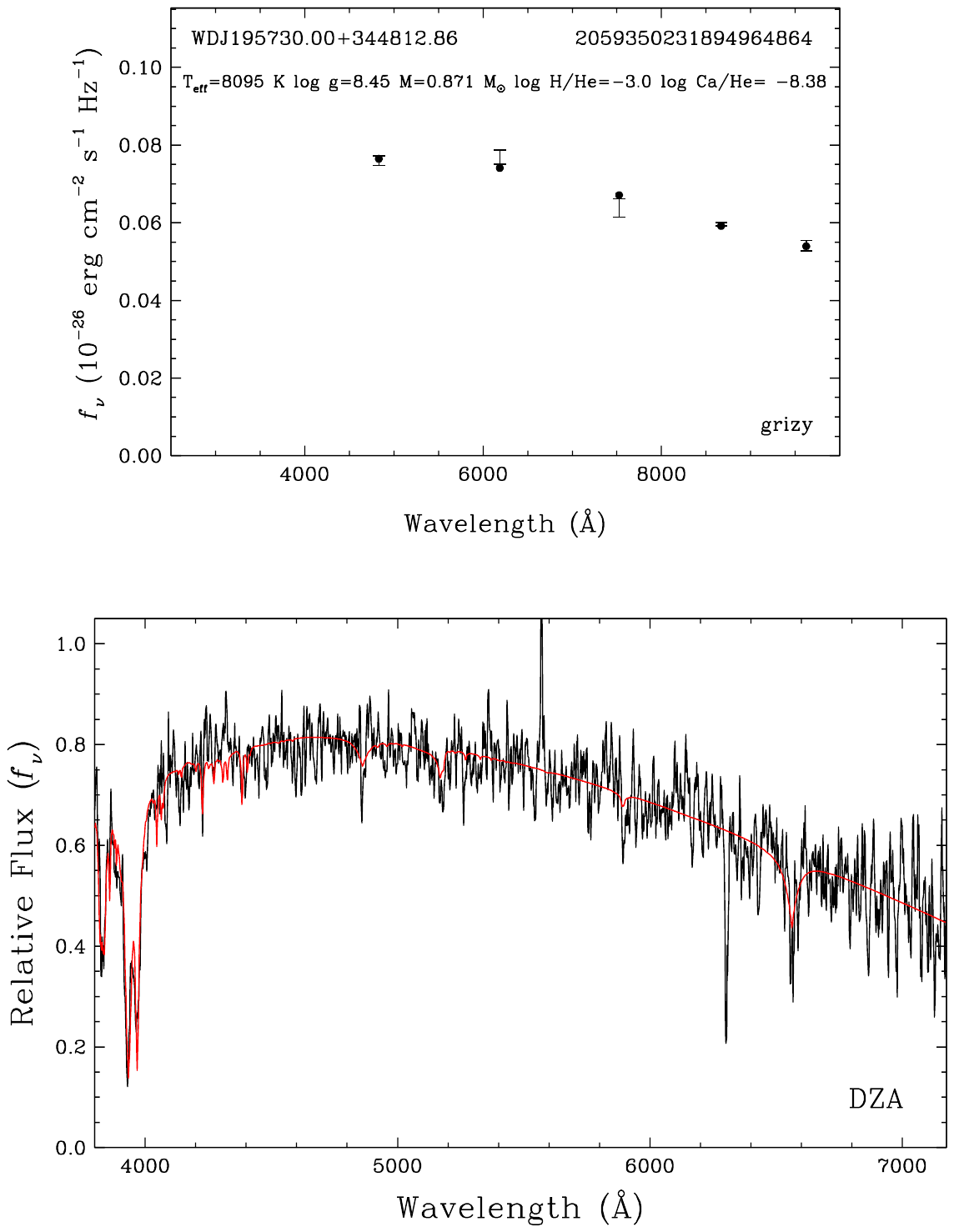}}
  \caption{Example of a white dwarf parameter fit with precise atmospheric models and combined photometry. This object, WD\,J195730.00+344812.86, is the same DZA as highlighted in Figure~\ref{fig:xpspec}. The complete figure set (65 images) is available in the online journal, and the derived atmospheric parameters are presented in Table~\ref{tab:dzs}.} \label{fig:fits}
\end{figure}

\begin{figure}[t]
  \centering
  {\includegraphics[width=.45\textwidth]{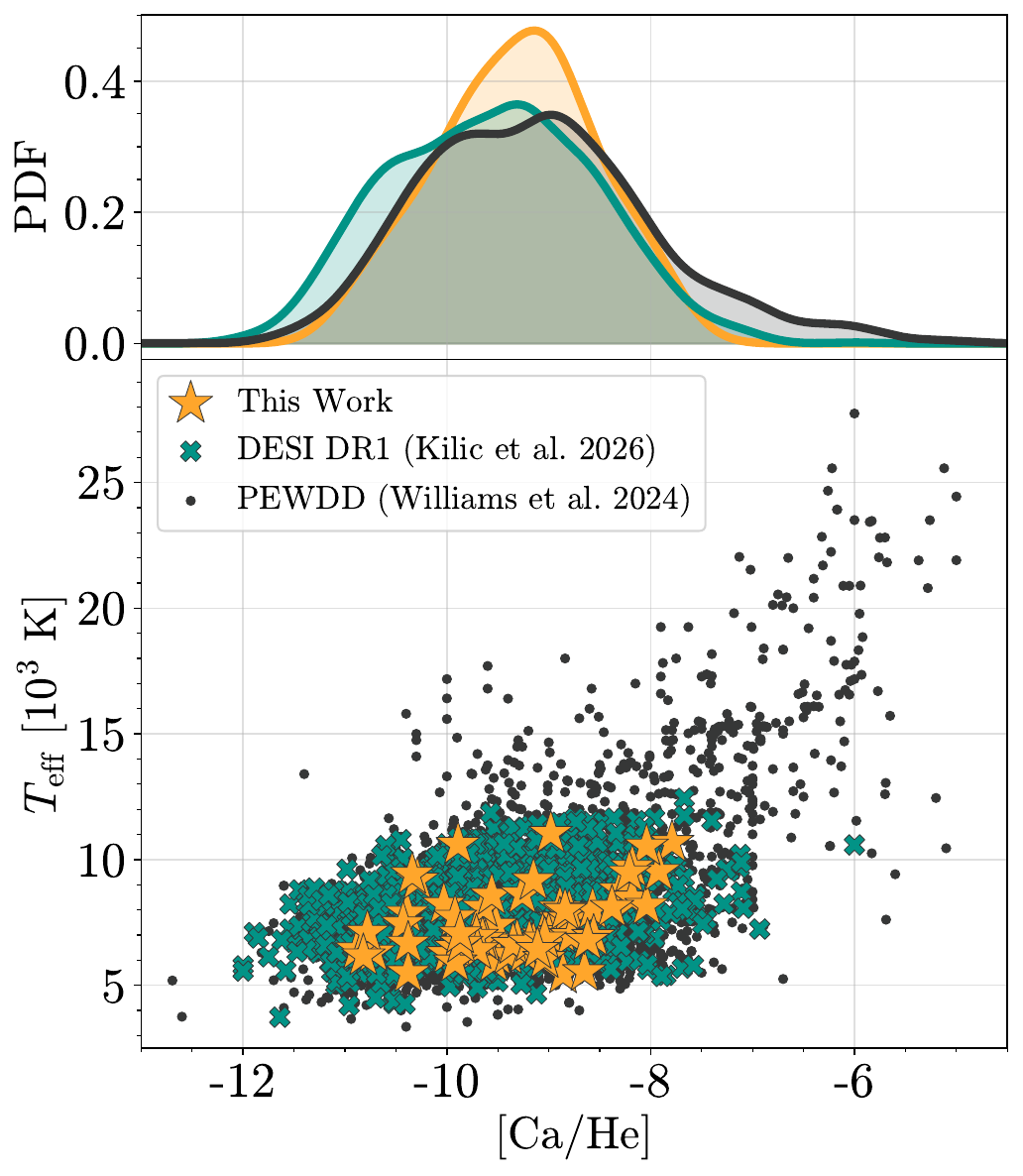}}
  \caption{\textit{Top:} Probability density function (PDF) of  measured Ca abundances, $\rm[Ca/He]$ (reported in Table~\ref{tab:dzs}), for the confirmed metal-polluted white dwarfs of our sample (orange), put in context with DZ white dwarfs in DESI DR1 which are fitted with the same technique (teal; \citealt{2026arXiv260625377K}), as well as all characterized white dwarfs in PEWDD (gray; \citealt{2024A&A...691A.352W}). While overall consistent with one another, centered around $\rm[Ca/He]\approx$-9.5, our sample shows a tighter distribution than the comparative samples. \textit{Bottom:} The same Ca abundances plotted with respect to effective temperature, $T_{\rm eff}$, which shows that the higher abundances reported in PEWDD are contributed from hotter white dwarfs. The difference in the lower limit of our measurements suggests that Gaia XP spectra are biased against classifying white dwarfs with low Ca abundances as DZ.} \label{fig:pdf_abundances}
\end{figure}

\section{Mass Measurements of Massive DZ Candidates}\label{sec:masses}

\subsection{Mass Fits with Precise DZ Models}

As described in Section~\ref{sec:sample}, we initially estimated candidate masses using the Gaia photometric fits assuming He-dominated atmospheres by \citet{2021MNRAS.508.3877G}. 

To obtain more accurate parameters, we perform dedicated detailed photometric spectral energy distribution (SED) fits for the confirmed metal-polluted sample following \citet{2026arXiv260625377K}. We combine available SDSS $u,g,r,i$ photometry \citep{2009ApJS..182..543A} with Pan-STARRS $grizy$ photometry \citep{2020ApJS..251....6M}, and SkyMapper $u,v,g,r,i,z$ photometry \citep{2024PASA...41...61O}, and Gaia DR3 parallaxes to constrain the effective temperature and the solid angle, $\pi(R/D)^2$, where $R$ is the stellar radius and $D$ is the Gaia-inferred distance. Given the precision of Gaia parallaxes, the solid angle directly constrains the stellar radius and hence the mass through white dwarf evolutionary models \citep{2020ApJ...901...93B}. For white dwarfs beyond 100\,pc, we apply reddening corrections using extinction estimates from \citet{2021MNRAS.508.3877G}. 
Observed magnitudes are converted to observed flux densities using standard AB zero points and compared to synthetic fluxes computed from model atmospheres with the appropriate chemical composition. We define a $\chi^2$ statistic over all available bandpasses, weighted by photometric uncertainties, and minimize it using a Levenberg-Marquardt least-squares algorithm \citep{1986nras.book.....P} to obtain best-fitting parameters.
For each object, we iterate between spectroscopic abundance estimates and photometric fits until convergence is achieved. The fitting is performed using extended versions of the atmospheric-model grid tailored to He-dominated metal-polluted DZ atmospheres \citep{2018ApJ...863..184B, 2018ApJ...867..161B} adapted as described in \citet{2026arXiv260625377K}. The confirmed DZ WD\,J012444.74$-$222906.88 was not included due to the unavailability of its spectrum online. Figure~\ref{fig:fits} shows a representative example of our fits, and Table~\ref{tab:dzs} lists all fitted parameters, $M_{\rm DZ}$, $T_{\rm eff,\,DZ}$, and $\log(g)_{\rm DZ}$, as well as the measured abundances $\rm[Ca/He]$, and $\rm[H/He]$.

We place the $\rm[Ca/He]$ abundances in context of the characterized metal-polluted white dwarf of the literature in Figure~\ref{fig:pdf_abundances}, which shows a probability density function for the confirmed DZ white dwarfs in our sample alongside the DESI DR1 DZ white dwarfs fitted in \citet{2026arXiv260625377K}, and the metal-polluted white dwarfs of PEWDD as of February 2026 \citep{2024A&A...691A.352W} in the top panel. The bottom panel of Figure~\ref{fig:pdf_abundances} highlights the \teff\, distribution of the compared populations as a function of $\rm[Ca/He]$ abundances. PEWDD contains hotter targets with high-resolution spectroscopic characterization, so our lower abundances are likely due to observational biases. When comparing with DESI white dwarfs which were fit using the same technique, our metal-polluted white dwarfs from Gaia XP spectra seem to be biased towards higher Ca abundances. Below an abundance of $\rm[Ca/He]<$$-$10.9, Gaia XP spectra may not be sufficient to detect the Ca~{\sc ii} features.

\begin{figure}[t]
  \centering
  {\includegraphics[width=.43\textwidth]{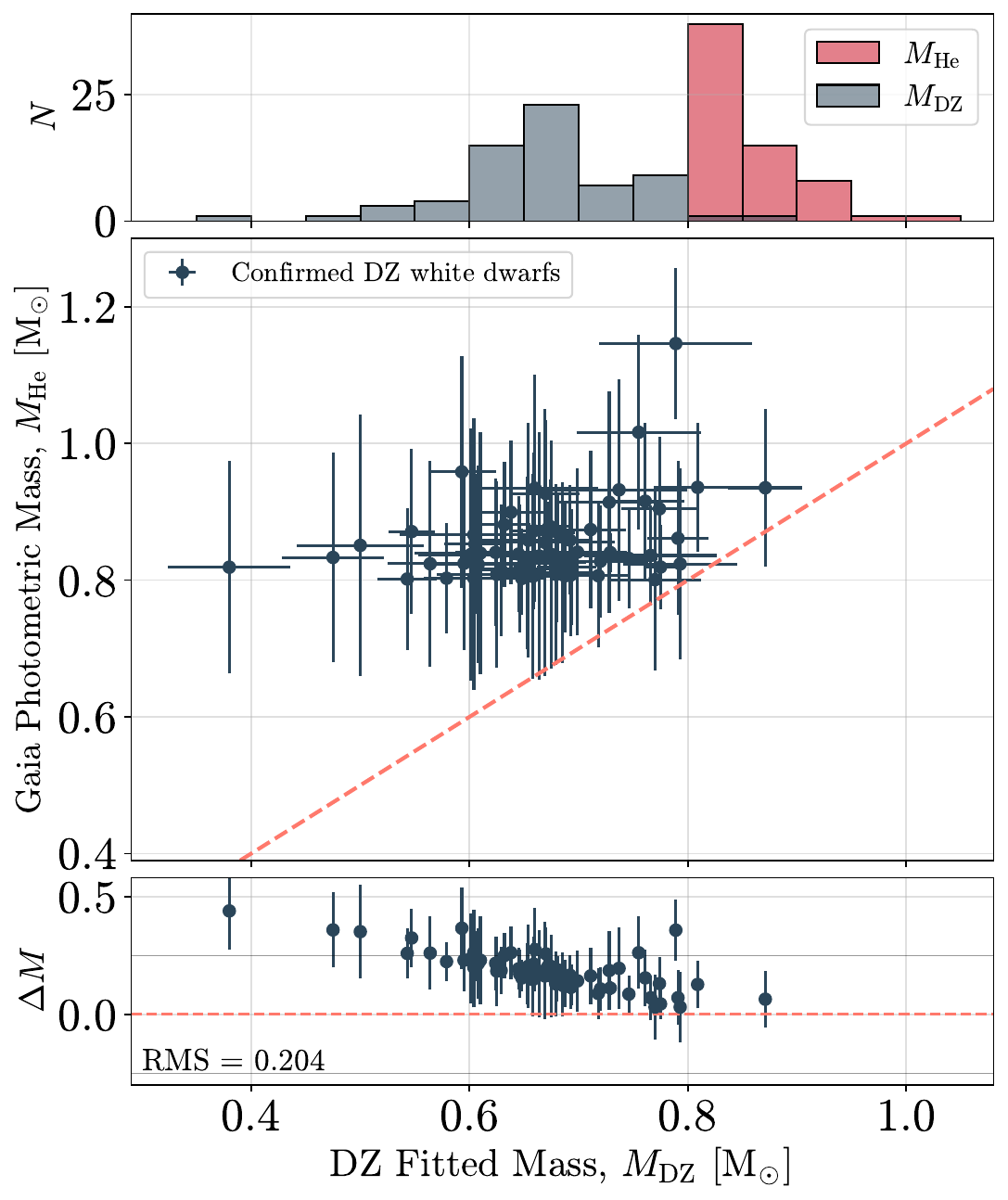}}
  \caption{Mass distribution of confirmed metal-polluted white dwarfs of our sample (top; gray histogram) compared to the Gaia photometric masses derived from pure-He atmospheric models (top; red histogram) The middle panel shows the relation between the two masses which lies above the $1{:}1$ relation (dashed red line). The bottom panel includes a residual plot of the mass difference. Gaia photometric masses are overestimated by a mean of +0.187\,\msun.} \label{fig:mass_comp}
\end{figure}

For the 65 fitted DZ white dwarfs, the masses derived from the DZ SED fits are systematically lower than the corresponding Gaia photometric masses, implying broad-band Gaia photometry alone systematically overestimates the masses of cool, metal-polluted white dwarfs. The shift in the mass of the confirmed DZ sample averages to $-0.187$ $\rm M_{\odot}$. 
Gaia photometric fits are derived from the position of each white dwarf in the Gaia color--magnitude diagram using the $G$, $G_{\rm BP}$, and $G_{\rm RP}$ magnitudes and pure-He atmospheric model grids. Photospheric metals change the continuum opacities, affecting the slope of the whole spectral energy distribution (SED), which leads to inaccurate \teff\, and \logg\, parameter estimates, and therefore inaccurate masses. 

Figure~\ref{fig:mass_comp} illustrates the difference in mass estimates between Gaia photometric fits and our precise DZ atmospheric fits with both spectra and multi-band photometry. The Gaia photometric masses are weakly correlated with the DZ fitted masses, with a residual root mean square (RMS) of 0.2. The revised mass distribution of the confirmed DZ sample is shown in gray and peaks at $0.67\,\rm M_{\odot}$. This distribution is broadly consistent with that of field white dwarfs, whose masses peak near $0.6\,\rm M_{\odot}$ and decline rapidly toward the high-mass tail above $0.8\,\rm M_{\odot}$ \citep{2016MNRAS.461.2100T, 2024MNRAS.527.8687O}. 

\begin{figure*}[t]
  \centering
  {\includegraphics[width=0.96\textwidth]{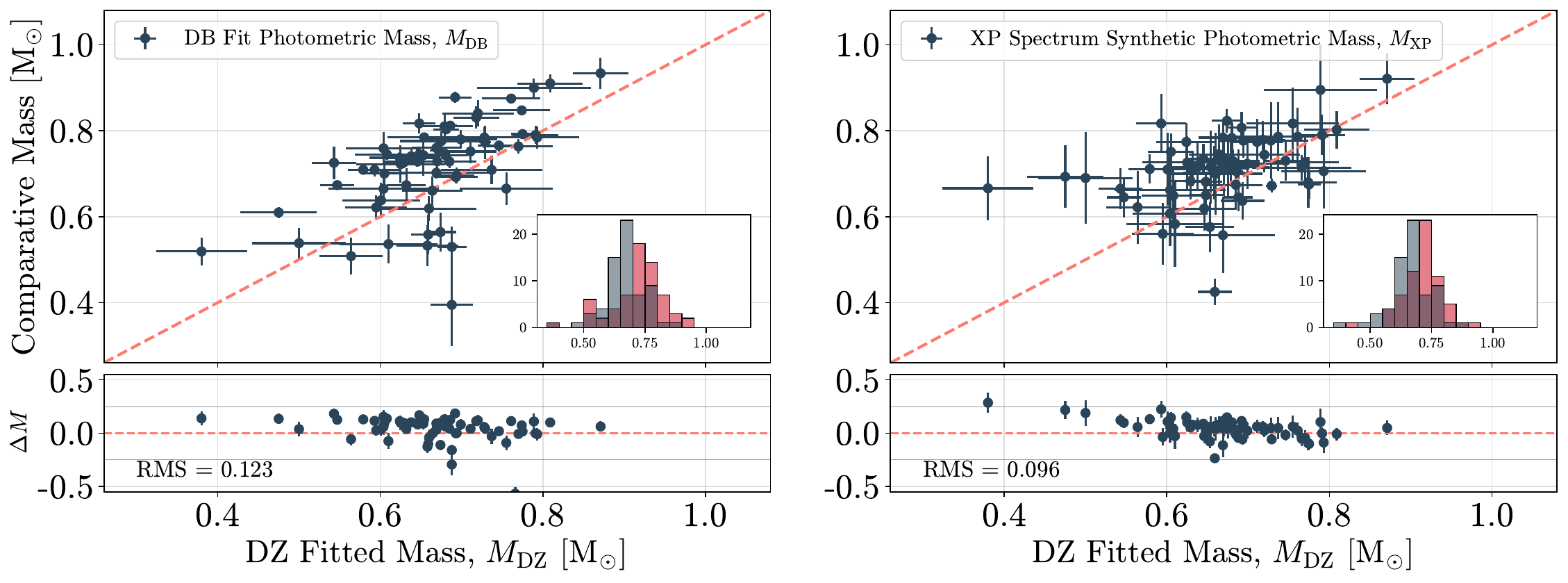}}
  \caption{Mass estimates for confirmed metal-polluted white dwarfs of our sample. Fitted mass estimates (y axis) are compared to the photometric-spectroscopic mass fits with DZ atmospheric models (x-axis, $M_{\rm DZ}$). The compared mass estimates are SDSS+Pan-STARRS+SkyMapper photometric fits with He-dominated models atmospheric models ($M_{\rm DB}$) and synthetic photometric fits to the XP spectra from \citet{2024A&A...682A...5V} ($M_{\rm XP}$), respectively. We show the $1{:}1$ relation for each panel, as well as an inset of the mass distribution for each population where $M_{\rm DZ}$ is shown in gray, and the compared mass distribution is shown in red. The residual mass shift of $M_{\rm compared} - M_{DZ}$ is plotted on the bottom panel. Gaia does not reproduce the masses of cool DZ white dwarfs, but widely sampled photometry SED fits and XP synthetic photometry fits have better agreement to the fitted DZ masses.} \label{fig:mass_comp_photometry}
\end{figure*}

\vspace{-3.5mm}
\subsection{Mass Fits in the Absence of Spectra}

The most accurate way to estimate masses for cool DZ white dwarfs is to use both spectra and photometry with the method described above. But in the absence of follow-up spectra, for DZ white dwarf candidates, we must rely on photometric fits. 

Masses derived directly from Gaia XP spectra, denoted here by $M_{\rm XP}$, use the photometric method from \citet{1997ApJS..108..339B} with DZ atmospheric models applied to standardized synthetic SDSS photometry from Gaia XP spectra, as detailed in \citet{2024A&A...682A...5V}. We find that this method estimates masses that are broadly consistent with the fitted DZ masses as shown on the right panel of Figure~\ref{fig:mass_comp_photometry} (residual RMS $<$0.1), and the offset is reduced to a mean overestimation of +0.042\msun. 

Without access to DZ atmospheric models, or if the spectral type of a cool white dwarf is unknown, photometric fits with narrower-band optical band photometry and pure-He atmospheric models estimate masses in closer agreement to the up-to-date DZ atmospheric mass fits than the Gaia photometric mass fits. Specifically, we use all available photometry from Pan-STARRS, SDSS, and SkyMapper to perform a pure-He atmosphere SED photometric fit, and derive masses denoted as $M_{\rm DB}$. When comparing $M_{\rm DB}$ to $M_{\rm DZ}$, the slope is still not consistent with unity, but the systematic compression of the Gaia photometric masses disappears, the scatter is reduced, and the recovered mass distribution broadly matches that obtained from the detailed DZ atmosphere fits (residual RMS $\approx$0.1) with a mean offset of +0.043\msun, similar to the Gaia XP mass estimates ($M_{\rm XP}$). This is shown on the left panel of Figure~\ref{fig:mass_comp_photometry}. We also perform the pure-He SED fit without the $u$-band photometry to try to establish what bandpass may be driving some of the overestimation. Without $u$-band photometry, we find a mean mass difference of +0.094\msun, or about half the offset found with pure-He Gaia photometry (+0.187\msun). Even without $u$-band coverage, once photometry is sufficiently well sampled across multiple bandpasses, SED fit techniques can recover reliable white dwarf masses, and provide substantial improvement over Gaia photometric masses.

From these methods, eight white dwarfs which were not spectroscopically followed-up within our study remain massive candidates, with both $M_{\rm XP} >$ 0.8\msun\, and $M_{\rm DB} >$ 0.8\msun, three of which have $P_{\rm DZ} > 0.65$. As such, these candidates are priority targets for future searches of massive, metal-polluted white dwarfs. Table~\ref{tab:sed_he} provides the He-atmospheric fitted parameters for the massive candidates that are priority follow-up targets, ordered by decreasing $P_{\rm DZ}$.

\begin{deluxetable*}{lccccr}[htbp!]
\tablecaption{Priority follow-up massive candidates with atmospheric parameters derived from photometric fits\label{tab:sed_he}}
\tablecolumns{6}
\tabletypesize{\footnotesize}

\tablehead{
    \colhead{Target Name} &
    \colhead{Gaia mass, $M_{\rm He}$} &
    \colhead{$G$ mag} &
    \colhead{He-model mass $M_{\rm DB}$} &
    \colhead{DZ-model mass $M_{\rm XP}$} &
    \colhead{$P_{\rm DZ}$}
}

\startdata
WD\,J070316.26$-$280326.71 & 0.994 (0.165) & 20.15 & 0.820 (0.023) & 0.817 (0.109) & 0.82 \\
WD\,J234737.30+611019.44 & 0.952 (0.122) & 19.71 & 0.846 (0.040) & 0.882 (0.072) & 0.82 \\
WD\,J192128.98+154504.92 & 0.928 (0.130) & 19.99 & 0.885 (0.014) & 0.887 (0.066) & 0.79 \\
WD\,J004644.27$-$331856.89 & 1.101 (0.105) & 19.56 & 1.031 (0.078) & 0.985 (0.056) & 0.62 \\
WD\,J002136.39+632408.57 & 0.864 (0.101) & 19.23 & 0.803 (0.035) & 0.801 (0.055) & 0.61 \\
WD\,J193132.65+055056.06 & 0.877 (0.136) & 19.46 & 0.875 (0.075) & 0.806 (0.096) & 0.60 \\
WD\,J025002.06$-$414220.40 & 0.893 (0.134) & 19.25 & 0.803 (0.055) & 0.900 (0.048) & 0.34 \\
WD\,J101541.53$-$200912.69 & 0.910 (0.125) & 19.41 & 0.966 (0.042) & 0.871 (0.054) & 0.23 \\
\enddata

\end{deluxetable*}



\section{The Search for Most Massive Metal-Polluted White Dwarfs with Gaia}\label{sec:discussion}


\subsection{New Massive DZ White Dwarfs}

When accounting for fitted DZ masses, $M_{\rm DZ}$ in our sample of confirmed metal-polluted white dwarfs, only two DZ white dwarfs remain with masses $M_{\rm DZ}> 0.8\,\rm M_{\odot}$. This result reinforces that genuinely massive metal-polluted white dwarfs are intrinsically rare \citep{2024ApJ...976..156O,2025MNRAS.539.2021C}. Most of the apparent massive DZ candidates selected from Gaia photometric parameters can be attributed to systematic mass overestimation rather than a large underlying population. 
The two new massive white dwarfs we present in this work are WD\,J035053.31+660433.38 (0.809~$\pm$~0.040\msun), and WD\,J195730.00+344812.86 (0.871~$\pm$~0.034\msun, highlighted in Figure~\ref{fig:xpspec} and Figure~\ref{fig:fits}). Both targets were observed with LDT/DeVeny. Notably, these white dwarfs have $P_{\rm DZ}$ of 0.64 and 0.55, respectively, which are both lower than the high-confidence threshold of $P_{\rm DZ} > 0.65$. These detections demonstrate that the Gaia XP catalog of \citet{2024A&A...682A...5V} provides an effective route toward identifying massive metal-polluted white dwarfs. However, our results also demonstrate that Gaia XP classification alone is insufficient for constructing samples of massive, metal-polluted white dwarfs. Accurate mass estimates from broadly sampled photometric SED fits or detailed atmospheric modeling remain necessary to distinguish genuinely massive systems from lower-mass DZ white dwarfs. Still, Gaia XP spectra classification is sensitive to the rare population of genuinely massive DZ white dwarfs.

Our search was not complete, as we only analyzed spectra for 97 of the 216 candidates of our sample. Table~\ref{tab:sed_he} provides eight additional massive candidates that are priority follow-up targets, based on their atmospheric parameters from SED fitting. Following our recovery rates, 82~$\pm$~5\% of targets with $P_{\rm DZ} > 0.65$ should be DZ white dwarfs (see Figure~\ref{fig:pdz_fraction}). 
From the XP spectra synthetic photometry mass estimates \citep{2024A&A...682A...5V}, there are an additional 10 targets with $M_{\rm XP}> 0.8\,\rm M_{\odot}$ that we did not observe, and are not included in Table~\ref{tab:sed_he}, only three of which have $P_{\rm DZ} > 0.65$ (WD\,J214858.46$-$444112.78, WD\,J185101.25$-$590724.76, and\\ WD\,J155812.29$-$652233.46), all visible from the Southern sky.  


\subsection{What is the Most Massive DZ Known?}

Massive DZ white dwarfs remain rare. The most massive metal-polluted white dwarf we identify is WD\,J195730.00+344812.86, with a mass of 0.871~$\pm$~0.034\msun. From initial-final-mass relations \citep{2018ApJ...866...21C, 2022ApJ...934..148H, 2024MNRAS.527.3602C}, we expect the progenitor of such a white dwarf to be 3.7~$\pm$~0.3\msun\, on the main sequence using the \texttt{wdwarfdate} package \citep{2022AJ....164...62K}. In order to comment on the upper limit of what stars can form and sustain planetary systems, we put this target in the context of known metal-polluted white dwarfs.

The Planetary Enriched White Dwarf Database (PEWDD; \citealt{2024A&A...691A.352W}) provides a comprehensive list of metal-polluted white dwarfs whose abundances have been measured. Of the nearly 1750 metal-polluted white dwarfs, PEWDD counts 134 objects with reported masses greater than 0.8\msun\,, of which just 19 white dwarfs exceed $M_{\rm WD} = 0.8$\msun\, with 3-$\sigma$ confidence. Even then, the lack of homogenized mass determinations prevents a consistent identification of the most massive system across this database. Specifically, some masses are determined from photometric fits while others are determined from spectroscopic fits. For many of these claimed massive white dwarfs, the Montreal White Dwarf Database reports various values statistically significantly inconsistent with one another. Of the 216 candidates in our sample, 15 appear in PEWDD, all with masses $M_{\rm DZ} <$0.8\msun. Without homogeneous mass determinations, it is difficult to assess whether these systems truly belong to the massive tail of the metal-polluted white dwarf population. 

Within this context, we seek to find known metal-polluted white dwarfs whose masses have been derived from precise DZ atmospheric models. In the 40\,pc white dwarf sample \citep{2023MNRAS.518.3055O, 2024MNRAS.527.8687O}, WD\,J085021.30$-$584806.21 (WD\,J0850$-$5848) stands out as one of the most massive metal-polluted white dwarfs ever published, with a well-characterized mass of 1.149~$\pm$~0.01\,M$_{\odot}$, and has a spectral type DZA. The inferred progenitor mass of WD\,J0850$-$5848 from initial-final-mass relation is 6.2~$\pm$~0.3 \msun\, using the same IFMR and package described earlier. 

Neither of the two new systems identified in this work exceeds the mass of WD\,J0850$-$5848, although they  expand the sample of reliable massive DZ white dwarfs and help refine the high-mass end of the distribution, strengthening the observational limits needed to characterize the rare remnant planetary systems hosted by the descendants of intermediate-mass stars.

Recently, white dwarfs in DESI DR1 have been analyzed with detailed atmospheric modeling \citep{2026arXiv260625377K}. Specifically, our DZ atmospheric fits utilize the same method, as described in Section~\ref{sec:masses}. Among the nearly 26,000 DESI DR1 spectra, 1,008 white dwarfs show photospheric metal pollution, classified with spectral types DZ, DZA, DAZ, DBAZ, or DZH. Interestingly again, only 12 have $M_{\rm DZ}> 0.8\,\rm M_{\odot}$ (and only one within 3-$\sigma$ confidence). Most noteworthy, two white dwarfs have masses above 1.05 \msun. WD\,J174308.26+251239.07 is a 1.064\,$\pm$\,0.101 \msun\, DBAZ with mixed H and He atmosphere and metal pollution, and a progenitor mass of 5.3\,$\pm$\,0.7\msun\, using the same IFMR and the \texttt{wdwarfdate} package. The sample of \citet{2026arXiv260625377K} also includes WD\,J141404.77+543825.59, which is a 1.053\,$\pm$\,0.166 \msun\, DAZ white dwarf, with progenitor mass of 3.8\,$\pm$\,1.5\msun. Five other targets have $M_{\rm DZ}> 0.9\,\rm M_{\odot}$, including two DQZ white dwarfs, one DZA white dwarf, and two DZ white dwarfs. DESI hence significantly expands the sample of massive, metal-polluted white dwarfs. 

\begin{figure*}[t]
  \centering
  {\includegraphics[width=\textwidth]{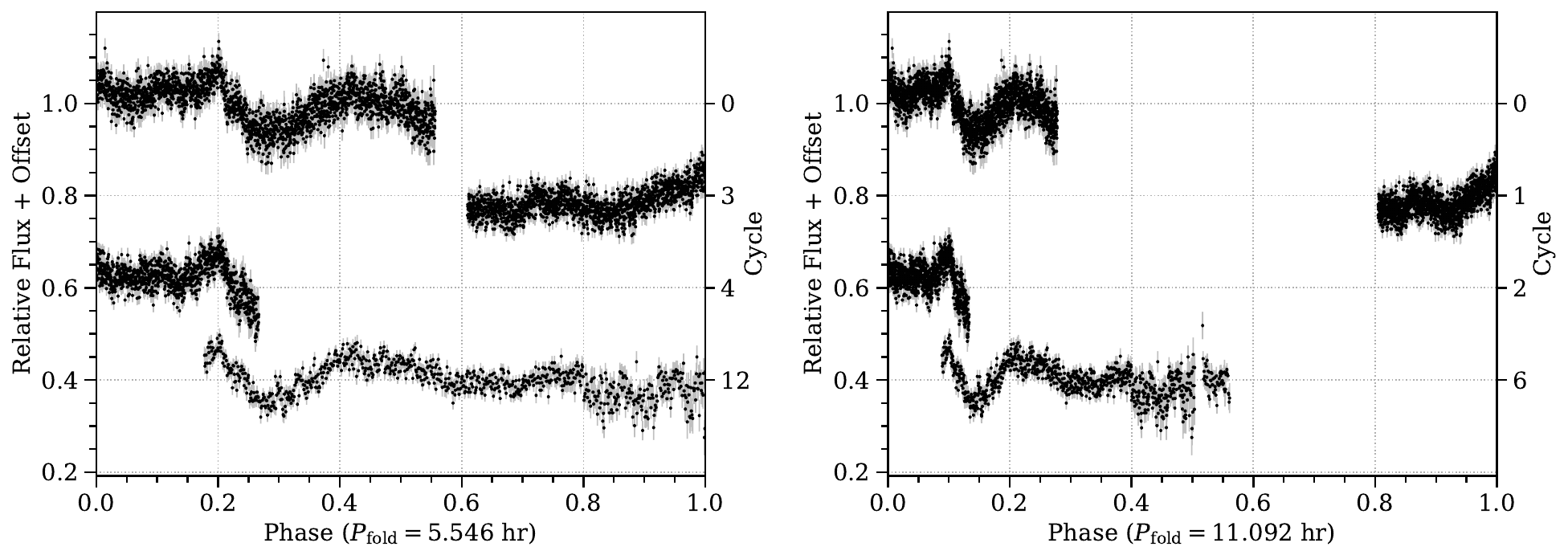}}
  \caption{Phase-folded light curves of SDSS\,J1624+5404, collected over the nights of 2025~May 25, 26, and 28 at McDonald Observatory. The left panel shows light curves folded on a period of 5.546\,hr, while the right panel is folded over 11.092\,hr. Both periods were significant in the ZTF light curve of the star, but we are unable to determine which period is the fundamental orbital period of the transiting debris system because of our incomplete phase coverage. The transits appear to continuously attenuate starlight across our whole observing baseline.} \label{fig:SDSSJ1624}
\end{figure*}


\subsection{New Transiting Debris System: SDSS\,J1624+5404}

WD\,J162458.18+540429.97 (SDSS\,J1624+5404) is a spectroscopically confirmed DZ with Gaia photometric mass $M = 0.82 \pm 0.06$\,M$_{\odot}$. Our LDT/DeVeny spectroscopy reveals strong Ca~{\sc ii}~H\,\&\,K absorption features. Our fit to this object returns $T_{\rm eff} = 10{,}530\,\pm$100\,K, and $M_{\rm DZ} = 0.775\pm0.014$\msun. 

We flagged SDSS\,J1624+5404 for photometric variability based on its anomalous Gaia $G$-band flux uncertainties \citep{2021ApJ...912..125G}. We searched our entire sample of massive DZ candidates for empirical photometric variability. Only three objects have {\sc varindex}$_{\rm eDR3} > 0$ within the sample: WD\,J032833.52$-$121945.27 (ZTF\,J0328$-$1219, a known DZ and transiting debris system \citealt{2021ApJ...917...41V}), WD\,J195730.00+344812.86 (a new DZ we confirm here), and SDSS\,J1624+5404, once selected as a candidate AM~CVn based on its photometric colors \citep{2013MNRAS.429.2143C}.

We queried $g$- and $r$-band light curves from the Zwicky Transient Facility (ZTF; \citealt{2019PASP..131a8002B,2019PASP..131a8003M}) Forced Photometry Service  \citep{2023arXiv230516279M} to probe for variability. We computed a PSD-normalized Lomb-Scargle periodogram \citep{1976Ap&SS..39..447L, 1982ApJ...263..835S} using the methods encoded by {\tt Astropy} \citep{astropy:2013,astropy:2018,astropy:2022} of the combined $g{+}r$ ZTF light curves, which exposed two peaks significant\footnote{\citet{2021ApJ...912..125G} found 5$\langle A \rangle$ a reliable proxy for statistically significant variability in ZTF DR3 light curves.} beyond $5 \langle A \rangle$ at 5.546\,hr and 11.092\,hr. We acquired follow-up, high-speed, time-series photometry with the blue-transmitting, red-cutoff {\em BG40} filter using the 2.1-meter Otto Struve Telescope at McDonald Observatory with the ProEM frame-transfer CCD mounted at Cassegrain focus. The observations were taken on the nights of 2025~May~25, 26, and 28 with exposure times ranging from $5-30$\,s. We reduced the images using standard Python-based methods, performed PSF photometry using the {\tt hipercam} pipeline \citep{Dhillon2021}, and extracted the optimal light curve using {\tt phot2lc} \citep{2023zndo...8169807V}, all following the methods described in \citet{2025ApJ...992..167G}. Because the transits continuously attenuate stellar flux, we do not normalize any of the light curves to a first- or second-order polynomial. 

Figure~\ref{fig:SDSSJ1624} shows our McDonald/ProEM light curves for this target. We observe SDSS\,J1624+5404 to be continuously occulted by planetary debris, never showing a clearly defined out-of-transit flux baseline, evoking the transit morphologies of ZTF\,J0328$-$1219 \citep{2021ApJ...917...41V} and WD\,1054$-$226 \citep{2022MNRAS.511.1647F}. The deepest, most readily identifiable transit appears to align in phase when folded on periods of 5.546 and 11.092\,hr, the two significant periods found from the ZTF light curve. While we are unable to rule as to which period is the fundamental, given the poor phase resolution of each orbit in our follow-up, we assert the detected variability is consistent with transiting planetary debris. Our LDT/DeVeny spectroscopy requires the magnetic field of SDSS\,J1624+5404 to be $B \lesssim 1\,$MG and we do not observe emission characteristic of a CV system. The sustained percent-level photometric variability is unlike the simpler sinusoidal modulations driven by magnetic surface spots rotating in and out of view \citep[e.g.,][]{2024ApJ...974..314O} or binarity  \citep[e.g.,][]{2024ApJ...967..166S}. Our discovery makes SDSS\,J1624+5404 the 21st published transiting debris system to date \citep{2025PASP..137g4202B, 2026arXiv260908923V}, of which only seven have measured orbital periods \citep{2015Natur.526..546V,2020ApJ...897..171V, 2021ApJ...917...41V, 2022MNRAS.511.1647F, 2025ApJ...980...56H, 2025ApJ...992..167G, 2026arXiv260908923V}. Whichever harmonic is the fundamental orbital period, a transit recurrence timescale of either 5.5 or 11.1\,hr would make SDSS\,J1624+5404 consistent with hosting a tidally disrupted asteroid \citep{2025MNRAS.537.2214L}.


\section{Conclusion}\label{sec:conclusion}

We obtained follow-up, low-resolution spectroscopy of 97 massive white dwarf candidates selected from Gaia XP spectra and confirm photospheric metals in 66 objects. Specifically, we found 82~$\pm$~5\% recovery of DZ white dwarfs in high confidence candidates of our sample (59/72 at $P_{\rm DZ} > 0.65$). The recovery efficiency increases with classifier confidence, reaching $94\pm3$\% for candidates with $P_{\rm DZ}>0.9$, demonstrating that Gaia XP spectra provide a reliable method for identifying cool, metal-polluted white dwarfs.

Using dedicated DZ atmospheric models, we show that masses inferred from broad-band Gaia photometry systematically overestimate the masses of cool, metal-polluted white dwarfs by an average of +0.187 \msun. After accounting for this bias, only two of the confirmed systems remain more massive than 0.8 \msun, WD\,J035053.31+660433.38 and WD\,J195730.00+344812.86.

Among false positive candidates, we most commonly identify continuum-dominated DC white dwarfs, but also describe outliers such as strongly magnetic white dwarfs, and a warm DQ white dwarf with high kinematics. We also identify an ultra-massive DBA white dwarf requiring higher-resolution spectroscopy to fully rule out the presence of photospheric metals. 

While our search did not reveal the most massive white dwarf able to sustain planetary systems, we discover a new transiting debris system at SDSS\,J1624+5404, with a transit orbital period of either 5.546 or 11.092\,hr.

Our results show that Gaia XP spectra efficiently identify cool, metal-polluted white dwarfs, though reliable identification of the most massive remnant planetary systems requires detailed atmospheric modeling and multi-band photometric SED fits. Continued spectroscopic follow-up of the remaining candidates, together with careful spectro-photometric mass estimates, will further refine the high-mass end of the metal-polluted white dwarf population and help constrain the most massive main-sequence stars that are able to form planetary systems.

\vspace{10mm}
\section{Acknowledgments}

The authors thank Olivier Vincent, Jamie Williams, Boris G\"{a}nsicke, and Laura Rogers for helpful discussions in the preparation of this manuscript, as well as the anonymous referee for their constructive comments. We are grateful for the observing contributions of Malia~L.~Kao, along with telescope operators at LDT, SOAR, and MagE. Specifically, we thank Jose Fernandez, Stephen Levine, Beck Maier, Ishara Nisley, Ben Shafransky, Cecilia Siqueiros, and Olivia Weiss at the Lowell Discovery Telescope, C\'esar Brice\~{n}o for the coordination of SOAR AEON observations, and John Kuehne, who provided instrumental support to our observing at McDonald Observatory.

L.\,B.\,O.\,R. is supported by the Future Investigators in NASA Earth and Space Science and Technology (FINESST F.5) program under Award No. 80NSSC24K1549. This material is based upon work supported by the National Aeronautics and Space Administration under Grant No. 80NSSC23K1068 issued through the Science Mission Directorate.
T.C. was supported by NASA through the NASA Hubble Fellowship grant HST-HF2-51527.001-A awarded by the Space Telescope Science Institute, which is operated by the Association of Universities for Research in Astronomy, Inc., for NASA, under contract NAS5-26555.

These results made use of the Lowell Discovery Telescope (LDT) at Lowell Observatory.  Lowell is a private, non-profit institution dedicated to astrophysical research and public appreciation of astronomy and operates the LDT in partnership with Boston University, the University of Maryland, the University of Toledo, Northern Arizona University and Yale University. 
This paper relied on observations obtained with SOAR programs \#866543, \#136460 and \#506309 through the Astronomical Event Observatory Network (AEON), a joint endeavor of the Las Cumbres Observatory and of NSF NOIRLab, which is managed by the Association of Universities for Research in Astronomy (AURA) under a cooperative agreement with the U.S. National Science Foundation.
This paper includes data gathered with the 6.5 meter Magellan Telescopes located at Las Campanas Observatory, Chile. This research uses services or data provided by the Astro Data Lab, which is part of the Community Science and Data Center (CSDC) Program of NSF NOIRLab. NOIRLab is operated by the Association of Universities for Research in Astronomy (AURA), Inc. under a cooperative agreement with the U.S. National Science Foundation. This work has made use of data from the European Space Agency (ESA) mission {\it Gaia} (\url{https://www.cosmos.esa.int/gaia}), processed by the {\it Gaia} Data Processing and Analysis Consortium (DPAC, \url{https://www.cosmos.esa.int/web/gaia/dpac/consortium}). Funding for the DPAC has been provided by national institutions, in particular the institutions participating in the {\it Gaia} Multilateral Agreement.
This research used data obtained with the Dark Energy Spectroscopic Instrument (DESI). DESI construction and operations is managed by the Lawrence Berkeley National Laboratory. This material is based upon work supported by the U.S. Department of Energy, Office of Science, Office of High-Energy Physics, under Contract No. DE–AC02–05CH11231, and by the National Energy Research Scientific Computing Center, a DOE Office of Science User Facility under the same contract. Additional support for DESI was provided by the U.S. National Science Foundation (NSF), Division of Astronomical Sciences under Contract No. AST-0950945 to the NSF’s National Optical-Infrared Astronomy Research Laboratory; the Science and Technology Facilities Council of the United Kingdom; the Gordon and Betty Moore Foundation; the Heising-Simons Foundation; the French Alternative Energies and Atomic Energy Commission (CEA); the National Council of Humanities, Science and Technology of Mexico (CONAHCYT); the Ministry of Science and Innovation of Spain (MICINN), and by the DESI Member Institutions: \url{www.desi.lbl.gov/collaborating-institutions}. The DESI collaboration is honored to be permitted to conduct scientific research on I’oligam Du’ag (Kitt Peak), a mountain with particular significance to the Tohono O’odham Nation. Any opinions, findings, and conclusions or recommendations expressed in this material are those of the author(s) and do not necessarily reflect the views of the U.S. National Science Foundation, the U.S. Department of Energy, or any of the listed funding agencies.

\vspace{2mm}
\facilities{DESI, Gaia, LDT (DeVeny), Magellan (MagE), SDSS, SOAR (Goodman), Struve (ProEM), Zwicky Transient Facility, Astro Data Lab \citep{2014SPIE.9149E..1TF, 2020A&C....3300411N}}
\software{Astropy \citep{astropy:2013, astropy:2018, astropy:2022}, astroquery \citep{2019AJ....157...98G}, ccdproc \citep{2017zndo...1069648C}, hipercam \citep{2021MNRAS.507..350D}, lmfit \citep{newville_2025_16175987}, matplotlib \citep{Hunter:2007}, numpy \citep{harris2020array}, pandas \citep{mckinney-proc-scipy-2010,the_pandas_dev_team_2024_13819579}, phot2lc \citep{2023zndo...8169807V}, PypeIt \citep{2020JOSS....5.2308P}, Pyriod \citep{2022ascl.soft07007B}, scipy \citep{2020SciPy-NMeth}, TOPCAT \citep{2011ascl.soft01010T}, wdwarfdate \citep{2022AJ....164...62K}}

\appendix

We include below Figure~\ref{fig:dqfit}, which shows the spectro-photometric fit to the warm DQ false positive white dwarf mentioned in Section~\ref{sec:false-pos}.

Table~\ref{tab:obs} describes our spectroscopic observations from LDT/DeVeny, SOAR/Goodman, and Magellan/MagE. For each setup, we include the target names, $G$ magnitudes, the date of observation, slit width, and exposures counts and duration. We also include signal-to-noise (S/N) per resolution element, measured in a 200\,\AA\ window centered near 4600\,\AA. The targets which have been processed with the DeVeny Pickup Noise Tool during the data reduction are denoted with $^*$. Finally, we report the spectral type classification for each object.

Figures~\ref{fig:tab:nonDZs-spec}, \ref{fig:tab:nonDZs-spec-2}, and \ref{fig:tab:nonDZs-spec-3} show all observed non-DZ targets, sorted by spectral type. We have not fit atmospheric parameters to these targets, but provide spectral type identifications and spectra sources (observed at LDT or SOAR). Figure~\ref{fig:tab:nonDZs-spec} includes two DA white dwarfs, with Balmer lines indicated in blue. Figure~\ref{fig:tab:nonDZs-spec-2} includes three strongly magnetic DH white dwarfs which do not show identifiable features and both have \teff\,$>$10,000\,K. Figure~\ref{fig:tab:nonDZs-spec-3} includes nine continuum-dominated white dwarfs with \teff\,$<$10,000\,K that we classify as DC white dwarfs. Further investigation is needed to rule out strong magnetism in these objects.

\renewcommand{\thefigure}{A.\arabic{figure}}
\setcounter{figure}{0}

\begin{figure*}[b]
\centering
  {\includegraphics[width=.95\textwidth]{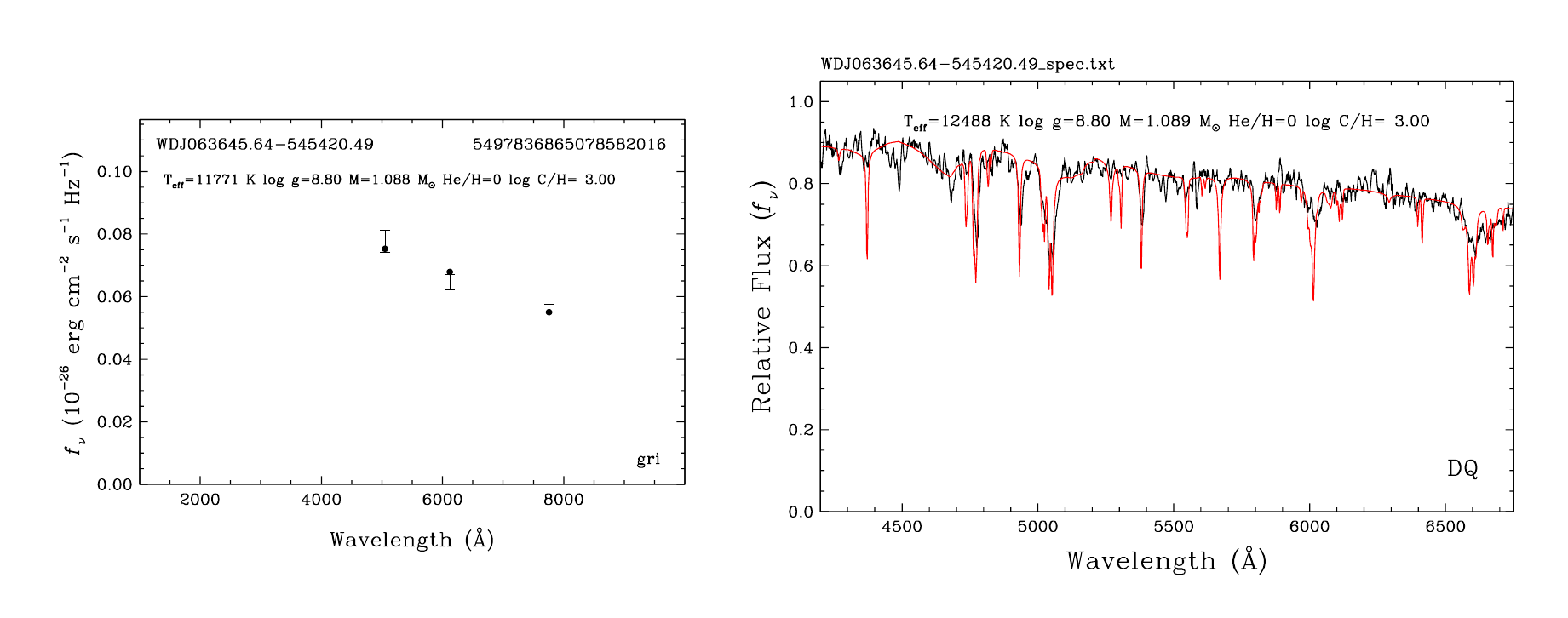}}
  \caption{Fit to massive DZ candidate reclassified as warm DQ, WD\,J063645.64$-$545420.49.} \label{fig:dqfit}
\end{figure*}

\begin{deluxetable*}{lcccccr}[htbp!]
\tablenum{A1}
\tablecaption{Spectroscopic observations sorted by target name \label{tab:obs}}
\tablecolumns{7}
\tabletypesize{\footnotesize}

\tablehead{\colhead{Target Name} &
    \colhead{$G$ mag} &
    \colhead{UT Date} &
    \colhead{Slit} &
    \colhead{Exposures (N $\times$ $t_{\mathrm{exp}}$)} &
    \colhead{S/N} &
    \colhead{Spec. Type}
    }
\startdata
\textbf{LDT/DeVeny DV2} & & & & & &\\
\hline
WDJ002617.27+001543.38 & 19.69 & 2024-11-02 & 3'' & 5 $\times$ 750\,s & 8.6 & DZ\\
WDJ003908.19+562002.49$^*$ & 18.94 & 2024-11-02 & 3'' & 5 $\times$ 600\,s & 11.0 & DZ\\
WDJ032833.52$-$121945.27 & 16.62 & 2024-09-13 & 1'' & 5 $\times$ 300\,s & 16.9 & DZ\\
WDJ035053.31+660433.38 & 19.38 & 2024-11-02 & 3'' & 4 $\times$ 600\,s & 8.6 & DZ\\
WDJ035345.55+401426.17 & 18.84 & 2024-09-05 & 3'' & 4 $\times$ 600\,s & 16.5 & DZ\\
WDJ050823.28+744430.34$^*$ & 19.04 & 2023-10-12 & 3'' & 6 $\times$ 900\,s & 7.6 & DC\\
WDJ055742.01+175256.25 & 18.90 & 2024-11-02 & 3'' & 4 $\times$ 600\,s & 13.0 & DZ\\
WDJ071608.21$-$145652.81 & 19.32 & 2024-11-02 & 3'' & 3 $\times$ 600\,s & 11.4 & DA\\
WDJ112206.33+842647.72 & 18.41 & 2025-04-28 & 3'' & 6 $\times$ 600\,s & 11.9 & DZ\\
WDJ112807.93+530625.77 & 18.94 & 2024-05-09 & 1'' & 5 $\times$ 600\,s & 9.8 & DZ\\
WDJ114305.14+080709.98 & 19.56 & 2024-05-10 & 1'' & 5 $\times$ 600\,s & 5.8 & DA\\
WDJ114843.35$-$152814.51 & 18.58 & 2026-05-09 & 1'' & 8 $\times$ 750\,s & 8.7 & DC\\
WDJ122314.50+800225.01 & 18.98 & 2026-05-09 & 3'' & 6 $\times$ 900\,s & 23.4 & DZ\\
WDJ130757.88+720836.58 & 18.97 & 2025-03-28 & 3'' & 4 $\times$ 600\,s & 8.5 & DC\\
WDJ152224.25+103951.36 & 19.10 & 2024-05-30 & 1'' & 4 $\times$ 600\,s & 11.4 & DZ\\
WDJ153746.70+301728.17 & 18.62 & 2024-06-05 & 3'' & 5 $\times$ 600\,s & 20.7 & DZ\\
WDJ155842.94$-$235841.58 & 18.41 & 2026-03-26 & 3'' & 4 $\times$ 600\,s & 15.4 & DC\\
WDJ155934.77+731357.85 & 18.62 & 2026-03-26 & 3'' & 6 $\times$ 900\,s & 15.8 & DC\\
WDJ162458.18+540429.97 & 17.91 & 2024-06-05 & 3'' & 4 $\times$ 600\,s & 20.3 & DZ\\
WDJ174213.13+155719.84 & 19.09 & 2024-09-05 & 3'' & 5 $\times$ 600\,s & 12.8 & DZ\\
WDJ181819.84+152434.51 & 18.35 & 2024-06-05 & 3'' & 5 $\times$ 600\,s & 23.9 & DC\\
WDJ185045.15$-$025939.58 & 18.81 & 2024-05-10 & 1'' & 4 $\times$ 600\,s & 14.4 & DZ\\
WDJ192632.10+560039.45 & 18.80 & 2024-06-05 & 3'' & 5 $\times$ 600\,s & 15.2 & DZ\\
WDJ195707.90+334441.41 & 19.74 & 2025-08-17 & 1'' & 10 $\times$ 900\,s & 10.2 & DZA\\
WDJ195730.00+344812.86 & 19.25 & 2023-10-12 & 3'' & 5 $\times$ 600\,s & 9.8 & DZA\\
WDJ195919.73+583634.93 & 19.10 & 2024-09-05 & 3'' & 4 $\times$ 600\,s & 13.1 & DH\\
WDJ195938.61$-$020829.97 & 18.61 & 2024-09-04 & 3'' & 4 $\times$ 600\,s & 16.2 & DZ\\
WDJ202617.90$-$002149.25 & 19.55 & 2024-09-05 & 3'' & 4 $\times$ 750\,s & 7.9 & DH\\
WDJ210119.94+812246.73 & 18.81 & 2024-11-02 & 3'' & 4 $\times$ 600\,s & 7.6 & DZ\\
WDJ211120.21$-$051032.16 & 18.14 & 2024-09-04 & 3'' & 4 $\times$ 600\,s & 19.4 & DZ\\
WDJ213909.61$-$070244.82 & 18.84 & 2024-09-04 & 3'' & 4 $\times$ 600\,s & 19.8 & DZ\\
WDJ215352.28+541408.86$^*$ & 19.29 & 2023-10-12 & 3'' & 6 $\times$ 600\,s & 12.4 & DC\\
WDJ223257.71+241303.62 & 18.53 & 2024-06-05 & 3'' & 7 $\times$ 600\,s & 18.8 & DZ\\
WDJ224511.73+684122.76 & 18.74 & 2024-09-05 & 3'' & 4 $\times$ 600\,s & 16.0 & DC\\
WDJ231747.18+370121.20 & 19.09 & 2024-09-04 & 3'' & 4 $\times$ 600\,s & 13.3 & DZ\\
\textbf{SOAR/Goodman 400-M1 Red Camera} & & & & & &\\
\hline
WDJ043019.38$-$294503.91 & 18.61 & 2025-10-10 & 1'' & 4 $\times$ 900\,s & 23.7 & DH\\
WDJ051753.43$-$434204.78 & 19.21 & 2025-11-08 & 1'' & 6 $\times$ 900\,s & 14.2 & DC\\
WDJ063645.64$-$545420.49 & 19.21 & 2026-04-26 & 1'' & 5 $\times$ 1200\,s & 9.2 & Warm DQ\\
\textbf{SOAR/Goodman 400-M1 Blue Camera}  & & & & & &\\
\hline
WDJ141030.66$-$315723.61 & 19.34 & 2026-05-09 & 1'' & 6 $\times$ 1200\,s & 13.3 & DC\\
WDJ150603.22$-$604016.49 & 19.29 & 2025-06-22 & 1'' & 4 $\times$ 1200\,s & 9.9 & DBA\\
\textbf{Magellan/MagE}  & & & & & &\\
\hline
WDJ150603.22$-$604016.49 & 19.29 & 2025-08-18 & 0.85'' & 4 $\times$ 1200\,s & 6.2 & DBA
\enddata
\tablecomments{S/N is measured per resolution element in a 200\,\AA\ window centered near 4600\,\AA. $^*$ Have been scrubbed with the DeVeny Pickup Noise Tool as described in Section~\ref{sec:observations}.}
\end{deluxetable*}

\begin{figure*}
\centering
  {\includegraphics[width=\textwidth]{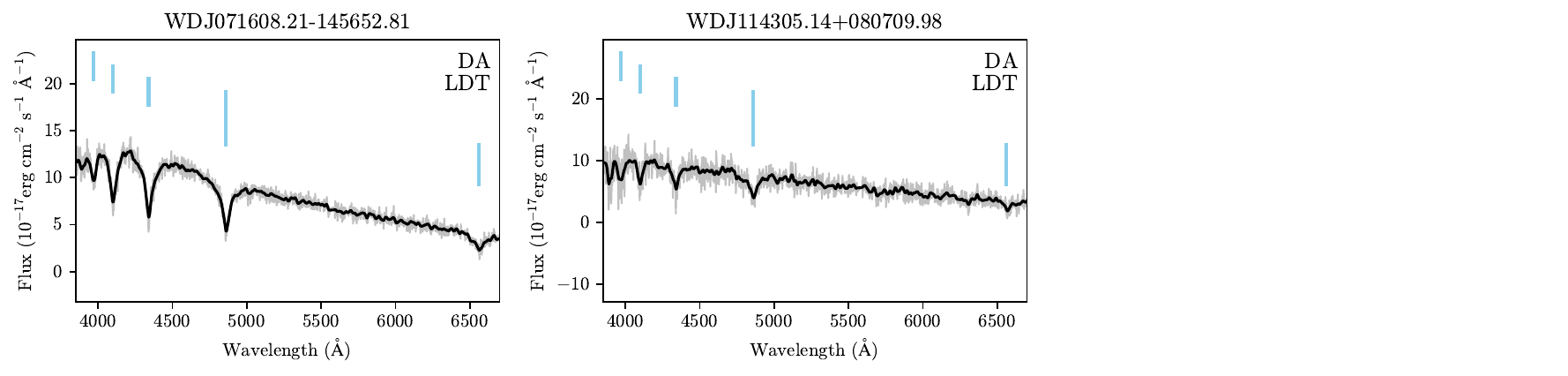}}
  \caption{Spectra of all new non-DZ white dwarfs in our sample reclassified as DA white dwarfs. We point out Balmer lines features with blue lines above each transition.} \label{fig:tab:nonDZs-spec}
\end{figure*}

\begin{figure*}
\centering
  {\includegraphics[width=\textwidth]{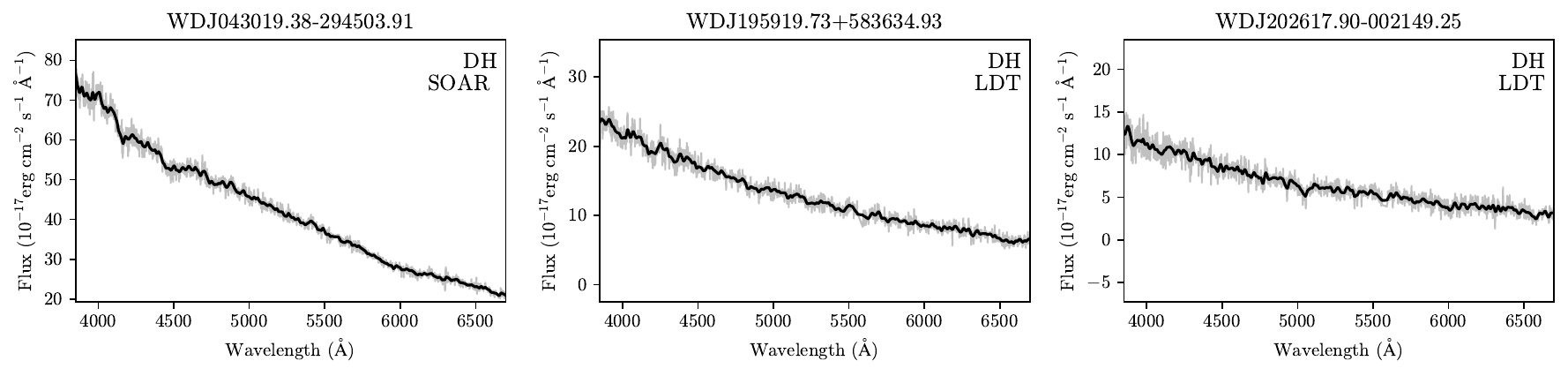}}
  \caption{Spectra of all new non-DZ white dwarfs in our sample reclassified as DH white dwarfs. These continuum white dwarfs are hot (\teff\,$>$10,000\,K) and should show absorption features from H or He unless they are strongly magnetic.} \label{fig:tab:nonDZs-spec-2}
\end{figure*}

\begin{figure*}
\centering
  {\includegraphics[width=\textwidth]{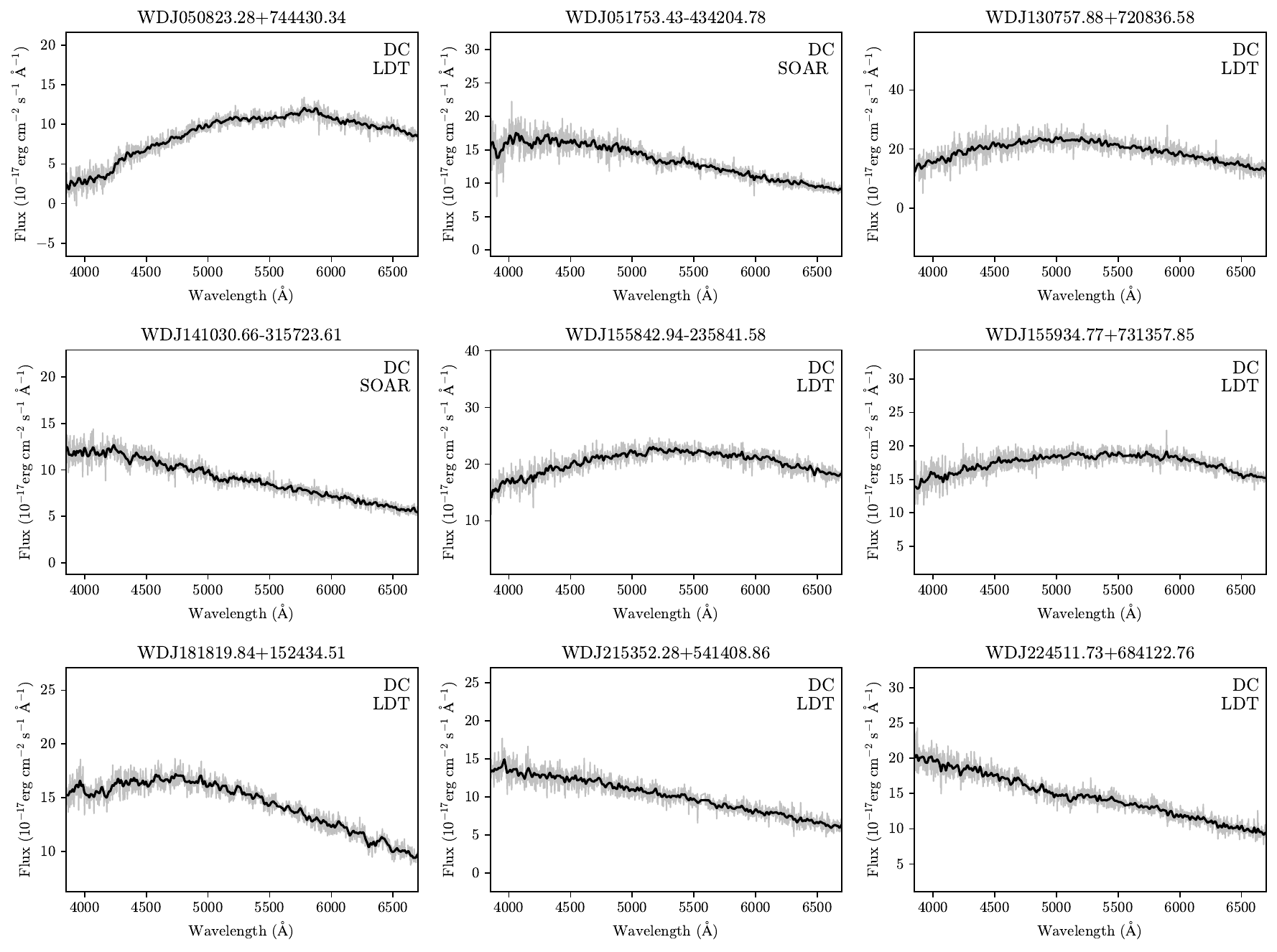}}
  \caption{Spectra of all new non-DZ white dwarfs in our sample reclassified as DC white dwarfs. Some wiggles in LDT spectra likely result from electronic interference from the spectrograph, even after using the DeVeny Pickup Noise Scrubber (see Section~\ref{sec:observations}).} \label{fig:tab:nonDZs-spec-3}
\end{figure*}


\clearpage

\bibliography{bibliography}{}
\bibliographystyle{aasjournal}
\end{CJK*}

\end{document}